\documentclass[10pt,journal,compsoc]{IEEEtran}
\usepackage{multirow}
\usepackage{caption}
\usepackage{subfigure}
\usepackage{booktabs}
\usepackage{graphicx}
\usepackage{amsmath, amsthm, amsfonts}
\usepackage{color}
\usepackage{cite}
\usepackage[justification=centering]{caption} 
\usepackage[ruled,linesnumbered]{algorithm2e}

\newtheorem{definition}{Definition}

\newcommand{\N}[1]{\textcolor[RGB]{0,0,0}{#1}}
\newcommand{\M}[1]{\textcolor[RGB]{0,0,0}{#1}}
\ifCLASSINFOpdf
\else

\fi
\begin{document}
%
\title{Fine-grained Spatio-Temporal Distribution Prediction of Mobile Content Delivery in 5G Ultra-Dense Networks}
\author{
    Shaoyuan~Huang,
    Heng~Zhang,
    \IEEEauthorblockN{Xiaofei~Wang\thanks{\IEEEauthorrefmark{1} Xiaofei Wang is corresponding author.}\IEEEauthorrefmark{1}} ,~\IEEEmembership{Senior~Member,~IEEE},
    Min~Chen,~\IEEEmembership{Fellow,~IEEE}\\
    Jianxin~Li,~\IEEEmembership{Member,~IEEE},
    and~Victor~C.~M.~Leung,~\IEEEmembership{Life~Fellow,~IEEE}
\IEEEcompsocitemizethanks{
\IEEEcompsocthanksitem  Shaoyuan  Huang, Heng Zhang and Xiaofei Wang are with  Tianjin  Key Laboratory of Advanced Networking, College of Intelligence and Computing, Tianjin University, Tianjin 300350, China. \protect\\
E-mail: \{hsy\_23, hengzhang, xiaofeiwang\}@tju.edu.cn
\IEEEcompsocthanksitem Min Chen is with the School of Computer Science and Technology, Huazhong University of Science and Technology, Wuhan, Hubei 430074, China. E-mail: minchen@ieee.org \protect 
\IEEEcompsocthanksitem Jianxin Li is with the School of Information Technology, Deakin University, Burwood, VIC 3220, Australia. E-mail: jianxin.li@deakin.edu.au \protect  
\IEEEcompsocthanksitem Victor C. M. Leung is with the College of Computer Science and Software Engineering, Shenzhen University, Shenzhen 518060, China E-mail: vleung@ieee.org \protect 
}
}

\markboth{IEEE TRANSACTIONS ON MOBILE COMPUTING}
{Shell \MakeLowercase{\textit{et al.}}: Bare Demo of IEEEtran.cls for Computer Society Journals}

\IEEEtitleabstractindextext{%
\begin{abstract}



The 5G networks have extensively promoted the growth of mobile users and novel applications, and with the skyrocketing user requests for a large amount of popular content, the consequent content delivery services (CDSs) have been bringing a heavy load to mobile service providers. As a key mission in intelligent networks management, understanding and predicting the distribution of CDSs benefits many tasks of modern network services such as resource provisioning and proactive content caching for content delivery networks. However, the revolutions in novel ubiquitous network architectures led by ultra-dense networks (UDNs) make the task extremely challenging. Specifically, conventional methods face the challenges of insufficient spatio precision, lacking generalizability, and complex multi-feature dependencies of user requests, making their effectiveness unreliable in CDSs prediction under 5G UDNs. In this paper, we propose to adopt a series of encoding and sampling methods to model CDSs of known and unknown areas at a tailored fine-grained level. Moreover, we design a spatio-temporal-social multi-feature extraction framework for CDSs hotspots prediction, in which a novel edge-enhanced graph convolution block is proposed to encode dynamic CDSs networks based on the social relationships and the spatio features. Besides, we introduce the Long-Short Term Memory (LSTM) to further capture the temporal dependency. Extensive performance evaluations with real-world measurement data collected in two mobile content applications demonstrate the effectiveness of our proposed solution, which can improve the prediction \M{area under the curve (AUC)} by 40.5\% compared to the state-of-the-art proposals at a spatio granularity of 76m, with up to 80\% of the unknown areas.
\end{abstract}





}

\maketitle

\IEEEdisplaynontitleabstractindextext

%
\IEEEpeerreviewmaketitle


\IEEEraisesectionheading{\section{Introduction}\label{sec:introduction}}
\IEEEPARstart{\N{W}}{\N{ith}} \N{the popularization of mobile networks, user requests for mobile application content \cite{9817385, 7839234} are growing explosively. When it comes to the 5G era, as the number of users and novel applications like Virtual Reality and cloud games soar, user content requests, and more importantly, the consequent content delivery services (CDSs), will account for a significant portion of mobile traffic in the networks \cite{7539325}.}



\N{More complex user requests and lower latency requirements place greater pressure on operators and mobile service providers. To address these challenges, significant efforts are devoted to optimize the mobile network. While at the architecture level, to meet the characteristics of proximity transmission due to the high-frequency band of 5G, ubiquitous networks are proposed to push the network resources (e.g., access points (APs)/content allocation/computing) to the edge of users to reduce the latency of CDSs. On the other hand, at the algorithm level, more network intelligence technologies are proposed to identify user requests and the corresponding CDSs automatically. Compared to reacting to users' requests passively, such network intelligence technology enables operators and mobile service providers to proactively improve their resource provision or cache strategies to achieve higher network throughput and lower service latency \cite{8865093}.}


\subsection{CDSs Prediction under 5G Networks}

\N{Revolutions in novel ubiquitous network architectures have brought paradigm-level changes to mobile networks. Specifically, the Ultra-Dense Networks (UDNs) empower 5G tremendous access capability, composed of extensive dense-deployed small cell base stations (SBSs). Additionally, with the assistance of Mobile Edge Computing (MEC), UDNs are capable of providing intelligent and efficient CDSs for user requests, which is implemented by SBSs edge servers. By contrast, the network intelligence technologies for CDSs prediction in 5G UDNs are still at an early stage and facing the following challenges:}


\N{\textbf{($i$) Insufficient spatio precision.} The APs in the ubiquitous networks are becoming extremely dense, making each AP serve a smaller range and the distribution of CDSs become spatio fine-grained. In UDNs, each SBSs serves user requests within 100m \cite{7476821}, while MEC's edge servers are also deployed within the nearest hop of the network from users \cite{8030322}. However, existing studies on the analysis and prediction of CDSs are scoped to regions or even entire cities \cite{10.1145/3209582.3209606, 8845204, article, wang2015characterizing}. These spatio coarse-grained approaches are not compatible with the service granularity of 5G UDNs and provide very limited guidance for APs under UDNs or MEC. To make network intelligence techniques compatible with the needs of the 5G network ecology, a method that can perform spatio fine-grained prediction of CDSs is necessary.}


\N{\textbf{($ii$) Lack of generalizability for predicting unknown areas.} \M{Compared to traditional networks, user mobility has a greater impact on CDSs in ubiquitous networks.} UDNs and MEC aim to provide closer communication and computational resources to hotspots of CDSs such as transportation stations, shopping malls and factories. However, such hotspots may migrate from old locations to adjacent areas as users move, requiring the methods that can accurately capture CDSs for these unknown areas (\emph{without historical data coverage}). However, existing studies mostly focus on predicting CDSs under a fixed area \cite{8117559, 8737488, 8057090} and cannot predict unknown areas with the possibility of CDSs appearing.}

\N{\textbf{($iii$) Complex multi-feature dependencies of user requests.} Extensive studies and dataset analysis have demonstrated that users' mobile content requests and the consequent CDSs have spatio, temporal and social dependencies. A comprehensive features capture model is needed to improve the CDSs prediction accuracy of network intelligence. While there are models that capture the temporal or spatio dependency of CDSs \cite{10.1145/3209582.3209606, 8292737, 7890496}, few studies consider more than two dependency features (especially social dependency, which has a significant impact on CDSs.)}



\N{To fill the gap between the network intelligence techniques and 5G ubiquitous networks, in this paper, we address the problem of spatio fine-grained prediction of mobile CDSs.}

\captionsetup[figure]{singlelinecheck=off, justification=justified}  
\begin{figure}
    \centering
    \includegraphics[width=8.5cm, height=4.867cm]{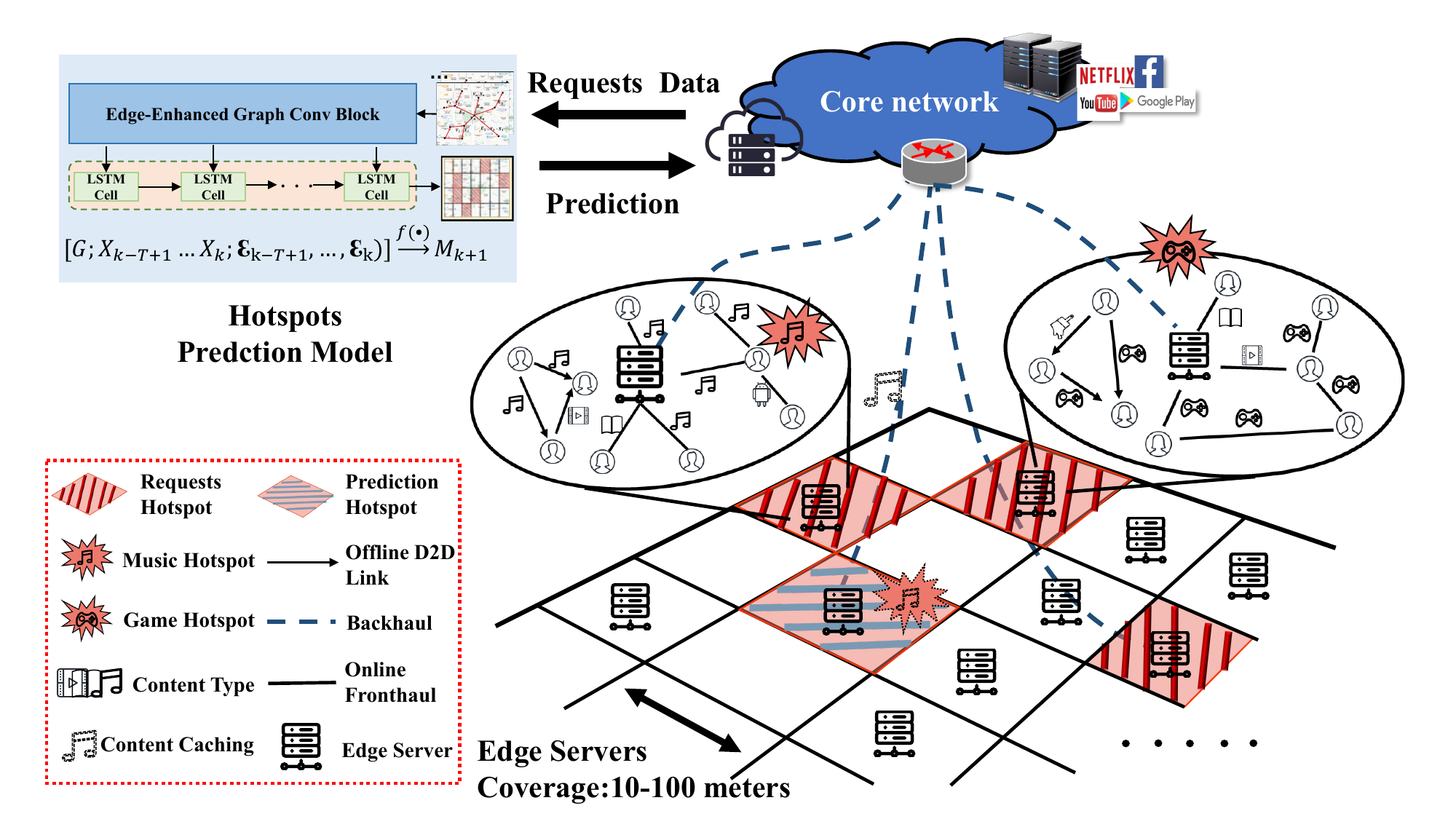}
    \caption{The Pipeline of the Designed Network Intelligence.}
    \label{fig:scene description}
\end{figure}
\clearcaptionsetup{figure}



\N{We use Fig. \ref{fig:scene description} to show the pipeline of our designed network intelligence. First, there will be a dense distribution of edge servers (or SBSs) in the cities covered by 5G UDNs. \M{These servers generally cover tens of meters and can handle and record a mass of mobile users' heterogeneous CDSs. The network intelligence deployed in the central cloud servers of the core network will then process the records data and train the prediction model that identifies how the spatio and temporal distribution of CDSs will evolve.}}

To make the problem more focused, we propose that the output of the prediction model is not just the location of CDSs but the \textbf{hotspots} (as defined in Section \ref{problem def}) for a specific category of content as shown in the red area in Fig. \ref{fig:scene description}. The hotspots reflect the areas with the most CDSs over time, and large-scale hotspots analysis can be used by the core network to guide SBSs optimization at various areas throughout the city.



\subsection{Paper Contributions}

In this paper, we propose a network intelligence oriented 5G UDNs. Specifically, we propose a spatio-temporal graphs modeling algorithm to model CDSs into CDSs networks at the specific spatio granularity. Besides, we propose a multi-feature extraction framework tailored to solve the spatio fine-grained hotspots prediction with high accuracy and spatio generalizability. We summarize our contributions into three aspects:


\N{($i$) \textbf{Spatio fine-grained encoding with high generalizability based on Geohash.} To refine the spatio granularity and model mobile CDSs of arbitrary areas (including known and adjacent unknown areas), we introduce the Geohash algorithm to encode the GPS information with specific granularity. After that, we propose a spatio-temporal graphs algorithm that can model CDSs networks for the CDSs records collected from mobile devices and embed the real-world spatio-temporal attributes.}

%
($ii$) \textbf{Spatio-temporal-social driven hotspots prediction model with edge enhanced block.} To maximize the predictor's accuracy, we analyze the main features affecting the distribution of CDSs from datasets. After that, we propose a hybrid model named Temporal-Edge Enhanced Graph Convolution Network (T-EEGCN) that simultaneously captures spatio, temporal, and social features to predict CDSs hotspots. The core lies in an elaborate Edge-Enhanced Graph Convolution Block (EEGCB), which is designed to encode the \textbf{dynamic} CDSs networks based on the extraction of \textbf{spatio-social features}. The EEGCB solves two major problems of incorporating real-world spatio features in virtual social networks and handling dynamic graphs (which is difficult for traditional GCNs). After that, we introduce LSTM to learn the CDSs networks encoding at different times and extract the \textbf{temporal dependency} of CDSs.



\N{($iii$) Extensive performance evaluations with a large number of measurements collected in two real-world CDSs applications demonstrates that our approach outperforms a broad benchmarks based on traditional and state-of-the-art prediction models. Besides, we perform ablation studies to analyze the role of different components of T-EEGCN and the dependencies of CDSs. Finally, we compared the performance of the proposed approach under different time scales settings to validate its temporal scalability.}

\M{The rest of this paper is organized as follows. In Section 2, we introduce the related works and highlight the novelty of our approaches. Formal definitions of key concepts and the problem are given in Section 3. Section 4 presents the proposed spatio-temporal graphs modeling and the T-EEGCN framework in detail. In Section 5, we evaluate the effectiveness of the solution via extensive experiments. We finally draw conclusions in Section 6.}

\section{Related Work}\label{sec:related work}
In this section, we review previous approaches regarding network intelligence for CDSs prediction and spatio-temporal prediction models.

\textbf{Network Intelligence for CDSs Prediction.} \M{\emph{1) Content popularity prediction.}} Collaborative filtering (CF) \cite{koren2022advances} is considered an effective way to uncover user demands for content and products. Previous CF algorithms only consider the user's interest in the content (ratings), which is effective but has significant limitations in accuracy. To improve the accuracy of CF, lots of studies try to incorporate spatio, temporal or social correlation considerations into the algorithm. Study \cite{lu2021time} explores collaborative filtering with temporal dynamics. Study \cite{Zeng2021PRRCUCAP} does collaborative filtering with user mobility and context. Nevertheless, the CF algorithms still face the problems of cold start, sparse data, and poor scalability, which means the CF cannot be applied to complex and large-scale mobile CDSs prediction.


\M{\emph{2) Traffic/capacity forecasting.}} There are two types of commonly used predicting methods: classic linear methods such as AutoRegressive Integrated Moving Average (ARIMA) \cite{wei2006time} and the other is machine learning based non-linear methods. In fact, non-linear methods gradually achieve better performance \cite{ABBASI202119, 10.1007/978-3-030-33778-0_11}. Study \cite{8057090} presents a hybrid deep learning model on the basis of autoencoder and Long-Short Term Memory networks (LSTM) to simultaneously capture the spatial and temporal dependency among different cells. \M{Besides, the city-scale wireless traffic predictions are also investigated in \cite{zhang2019deep, zhou2022large, 9184280}, in which the authors introduce novel prediction frameworks by modeling spatio-temporal dependency over cross-domain datasets.} In the latest study \cite{zhang2021dual}, the authors propose a novel wireless traffic prediction framework, by which a high-quality prediction model based on dual attention scheme is trained collaboratively by multiple edge clients.

\N{Regardless of forms, a major drawback of existing methods is that they can only predict CDSs in fixed areas and cannot address the new CDSs in unknown areas caused by user mobility. In addition, these methods are spatio coarse-grained (city-scale), making it difficult to predict CDSs in specific fine-grained areas.}


\textbf{Spatio-temporal Prediction Methods.} \emph{\M{1) Spatio-temporal analysis and feature extraction models.}} From the spatio domain, \cite{8845204} analyzes the spatio importance of telecommunication hotspots. The authors model the hotspots as nodes in a graph and then apply node centrality metrics that quantify the importance of each node. From the time domain, study \cite{shafiq2011characterizing} finds the traffic dynamics of CDSs can be well captured by Markov models. Realizing the spatio-temporal dependency of mobile CDSs, traditional methods \cite{6757900, 1203886} and machine learning models for spatio-temporal features extraction have been investigated in order to achieve accurate prediction of CDSs. \M{\cite{DUAN2022109156} propose a novel deep generative adversarial network to address the crowdsourcing-based urban cellular traffic prediction. In \cite{9751165}, the authors introduce a multivariate and spatio-temporal approach to predict cellular traffic, which considers cell features, peak hours and handover traffic.}

\N{In the latest studies, more novel deep learning based spatio-temporal predictors (e.g., Transformer) are proposed for user traffic prediction \cite{9112663}, passenger demand prediction \cite{9127090}, user mobility prediction \cite{8903457, 9165195}, and other fields.}

\emph{\M{2) Graph Convolution Networks (GCNs) for spatio-temporal prediction.}} In recent studies, GCNs \cite{kipf2017semisupervised} have been widely used in the spatio-temporal modeling of graph structure due to their capability of handling Non-Euclidean data. \cite{8117559} propose a graph-based deep learning approach to model in-tower and inter-tower traffic and learn long-distance spatio correlations for mobile traffic prediction. Both \cite{Bing2018Spatio} and \cite{ijcai2020-326} propose a convolutional neural network fusing GCN and GLU to capture spatio features of the roads network and temporal features in the transportation traffic domain.

Nonetheless, traditional GCNs assume a fixed structure among networks (e.g., BSs and roads networks), so they compute graph structure features once and set them as constant coefficients. In this study, the CDSs records of different time slots are modeled as a series of dynamic graphs (with different nodes and graph structures), so static graph convolution methods do not apply to our scenario.


All aforementioned studies mainly focus on spatio-temporal features but ignore the social relations among users. Since users share more frequently in current mobile applications, recent studies have found that social relations have a significant impact on mobile content requests \cite{qiu2018deepinf,2019Social2, Liu2019}. In this study, we exploit graph convolution to model user social relations. Moreover, the exploitation of edge features and novel information propagation/aggregation methods are implemented for capturing spatio dependency and addressing dynamic graphs, respectively.






\section{PRELIMINARIES}\label{problem def}
We proceed to introduce the background settings and formalize the hotspots prediction problem. Frequently used notations are summarized in Table \ref{notation}.

\begin{table}[htbp!]
\normalsize
\centering
\caption{Summary of Notations}
\label{notation}
\begin{tabular}{p{1.5cm}p{6.5cm}}
\toprule
\textbf{Notation} & \textbf{Description} \\ \midrule
  $R$  & The spatio granularity.         \\
  $I$  & The CDSs intensity, which is set to the number of contents requests.\\
  $M$  & The hotspots matrix. \\
  $\tau$  & Hotspots threshold. \\
  $\mathcal{G}^i_j$ & The CDSs network of area $i$ at time $j$. \\
  $G^i$ & The series of CDSs networks of area $i$. \\
  $X$ & The node features.\\ 
  $\mathcal{P}$ & The dimension of node features. \\
  $T$ & The size of time windows of $G$. \\
  $V$ & Node set of $\mathcal{G}$. \\
  $E$ & Edge set of $\mathcal{G}$. \\
  $\mathcal{E}$ & The edge features.\\ \midrule
  $geo_i^j$ & The $j$ bits Geohash encoding of node $i$. \\
  $a_{ij}$ & The edge weights of $edge_{ij}$. \\
  $h_i^l$ & The feature embedding of node $i$ in layer $l$. \\
  $m_{ij}$ & The information propagation to node $i$ from node $j$. \\
  $\mathcal{V}$ & The CDSs network embedding. \\
  $O$ & The output of LSTM. \\ 
  \bottomrule
\end{tabular}
\end{table}

\subsection{Settings}
\begin{definition}[\textbf{Spatio Fine-grained}]
\N{The spatio granularity $R$ refers to the radius of the predicted unit area. In the 5G UDNs architecture, the range served by any edge server or cell is greatly reduced, which requires the spatio granularity of the model's prediction to be changed from the previous coarse granularity (city-level) to fine granularity ($R<=100m$ \cite{7476821}).}


\end{definition}

\begin{definition}[\textbf{Unknown Areas}]
\N{Unknown areas are the areas with no CDSs appearing in the historical data}
\end{definition}

\N{We call the areas with CDSs appearing in the historical data known areas, and we predict the volume of CDSs in these areas and determine whether they become hotspots. In addition, we concern all unknown areas adjacent to known areas. Though no historical CDSs occur in these areas, we argue that they also have the potential to become hotspots in the foreseeable future due to user mobility. The unknown areas prediction is particularly important for service scheduling and device deployment in 5G UDNs.}

\begin{definition}[\textbf{Hotspots}]
Hotspots are areas with very high number of CDSs relative to others.
\end{definition}
To calculate the hotspots in all areas, we propose the hotspots delineation method based on the overall threshold with reference to \cite{8845204}. Specifically, if the number of CDSs of an area are greater than the over all threshold $\tau$, we set the area as a hotspot. The threshold $\tau$ is calculated as follows:
\begin{equation}
   \tau = \frac{1}{N} \cdot \sum_{a=1}^N I_a + (Max(\vec{I}) - \frac{1}{N} \cdot \sum_{a=1}^N I_a) \cdot P 
\end{equation}where $N$ is the number of areas, $I_a$ is the num of CDSs in area $a$, and $P$ is the parameter to determine the cutoff threshold.
\begin{definition}[\textbf{Hotspots Matrix}]
A hotspots matrix $M$ is a 0/1 matrix which represents the distribution of hotspots and non-hotspots in a large space with $N$ areas.
\end{definition} 
\begin{definition}[\textbf{CDSs Network}]
CDSs network $\mathcal{G}$ is the social network built from the CDSs records. Each node $v$ in CDSs networks represents once \textbf{CDSs}, e.g., video streaming  delivery. Each edge $e$ represents the social relation between users of twice CDSs, i.e., the two users have a friendly relationship.
\end{definition} 

\subsection{Problem Statement}
We use a series of snapshots $G=\{\mathcal{G}_{\kappa-T+1},...,\mathcal{G}_\kappa\}$ to represent different CDSs networks of $T$ time windows. Each snapshot $\mathcal{G}_t=(V_t,E_t)$ is a directed graph with different node set $V$ and edge set $E$. We use $X \in \mathbb{R}^{T \times |V_t| \times \mathcal{P}}$ as node features and $\mathcal{P}$ as the feature dimension (will be discussed in Section \ref{data_pro}). Besides, we use $\mathcal{E} \in \mathbb{R}^{T \times |V_t| \times |V_t|}$ as edge features. \M{The hotspots prediction problem can be formally described below: 
\begin{equation}\label{eq:map_f}
    [G;X;\mathcal{E}] \xrightarrow{f(\cdot)} \hat{M}_{\kappa+1}
\end{equation}where $f(\cdot)$ is the model that we need to construct. We can summarize the hotspots prediction task to a classification problem based on sequential graph data and the problem solving process can be formalized as follows.}
\M{
\begin{align}
\hat{M}_{\kappa+1}&=\underset{M_{\kappa+1}}{\arg \max}\ log P\left(M_{\kappa+1} \mid G, X, \mathcal{E}\right) \\
\text {s.t.} &\forall a \in M: R(a)<=100m \\
& \left\|G\right\|=T
\end{align}where the objective of CDSs hotspots prediction problem is to find the most likelihood CDSs hotspots matrix $\hat{M}_{\kappa+1}$ according to the history CDSs networks $G$, node features $X$, and edge features $\mathcal{E}$.}

\M{There are two main constraints to limit the problem sovling. In the first constraint (eq. 4), we limit the spatio granularity $R$ of every area $a$ to less than 100m, since we should guarantee that the model can accurately distinguish the distribution of CDSs in different fine-grained areas over a wide range of regions. Besides, we limit the periods length of the historical data to $T$ since we need to trade off the impact of historical information on the prediction and the temporal granularity.}

\begin{table}
\caption{$R$ of Different Geohash Encoding Lengths.}
\label{spatio granularity}
\normalsize
\centering
\begin{tabular}{cccc}
\hline
\textbf{Geohash length} &\textbf{lat bits} &\textbf{lng bits} &\textbf{R(m)}  \\ \hline
5 & 12 & 13 & 2400 \\ 
6 & 15 & 15 & 610 \\
7 & 17 & 18 & 76 \\
8 & 20 & 20 & 19 \\ \hline
\end{tabular}
\end{table}

\section{Fine-grained Hotspots Prediction}\label{npd}
\begin{figure}[bp]
    \centering
    \includegraphics[width=\columnwidth]{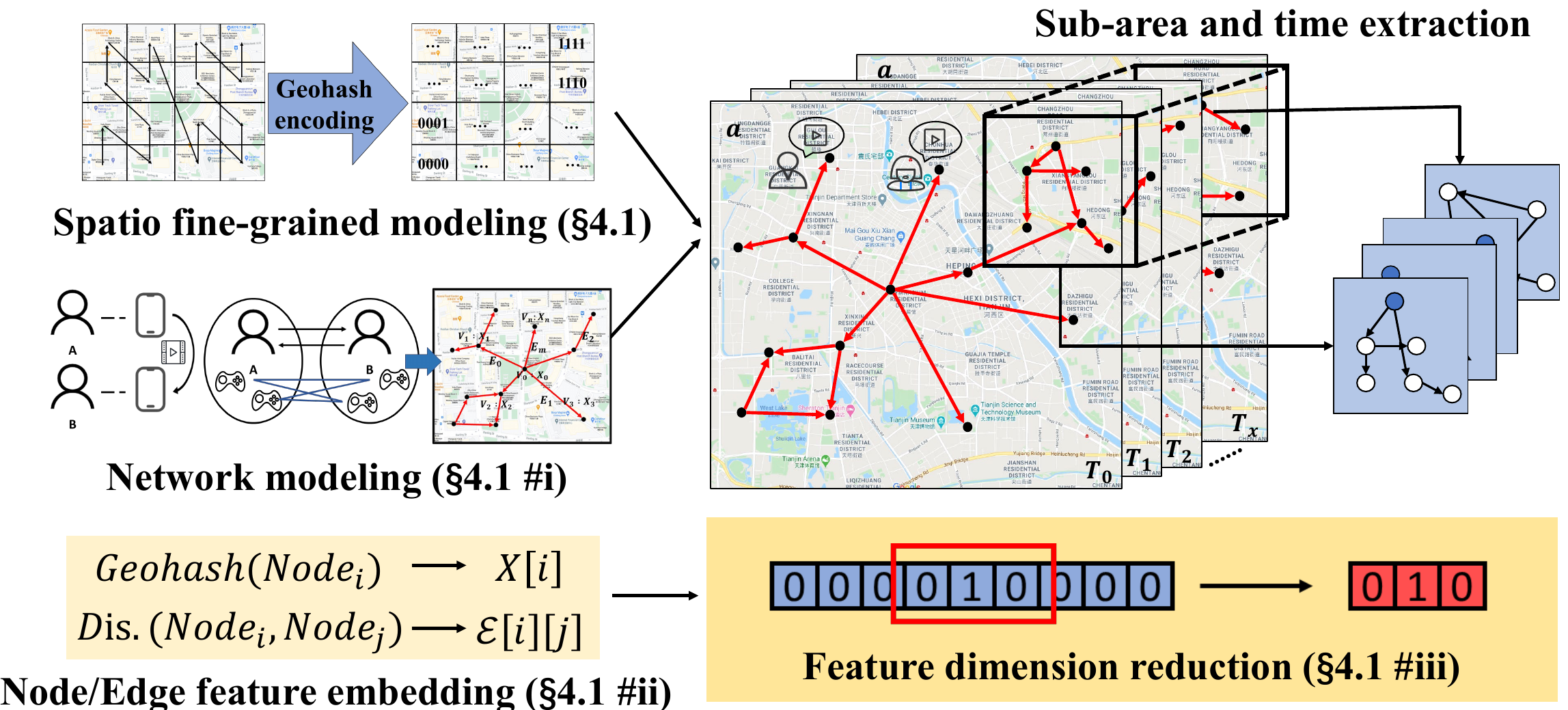}
    \caption{Spatio-Temporal Graphs Modeling}
    \label{fig:st-graph-modeling}
\end{figure}

\N{We proceed to introduce an end-to-end deep learning model, called temporal-edge enhanced graph convolution network (T-EEGCN), to predict spatio fine-grained CDSs hotspots. First, we introduce a spatio information encoding method, based on which we propose a spatio-temporal subgraph sampling algorithm, the complete encoding and algorithm process is shown in Fig. \ref{fig:st-graph-modeling}. Then, we present the details of T-EEGCN.}







\subsection{Spatio-Temporal Graphs Modeling}

\N{The GPS of CDSs are key information to model the space, which can be automatically obtained by the mobile terminal when users make requests. However, since the GPS encoding is global-level, its spatio granularity is much larger than our problem concerns, so we need a method to further refine the encoding of spatio information to help distinguish between fine-grained areas.}

\N{In addition, we need to access the adjacency between areas and equally encode these areas, which is a prerequisite for predicting unknown areas.}

We introduce the Geohash encoding algorithm \footnote{http://geohash.org/site/tips.html} to encode GPS information of CDSs into short strings of letters and digits. Geohash is a hierarchical spatio data structure that subdivides space into buckets of grid shapes. It allows areas of arbitrary precision to be encoded by increasing the encoding length. Considering Geohash's accuracy and the requirement of spatio fine-grained (as shown in the Table \ref{spatio granularity}), we set the encoding length of Geohash to \textbf{7bits}, i.e., the concerned spatio granularity $R = \textbf{76m}$.

\N{Another feature of Geohash is the hierarchical encoding, which guarantees that the longer a shared prefix between two geohashes is, the spatio closer they are together. Therefore, we can obtain the encoding of adjacent areas of known areas based on the prefixes of existing codes. Although Geohash encodes grid shape areas, we argue that in our modeling process, Geohash can be replaced by other shape geocoding algorithms as long as it can encode areas of arbitrary precision and encode adjacent areas.}

After determining the spatio granularity and the encoding of adjacent areas, we propose the spatio-temporal graphs modeling, i.e., the algorithm \ref{alg:Spatio-Temporal Graphs} to transform CDSs records into CDSs networks. The spatio-temporal graphs modeling algorithm achieves the following three main goals:

\begin{algorithm}
\caption{Spatio-Temporal Graphs Modeling}
\label{alg:Spatio-Temporal Graphs}
\KwIn{CDSs Records Set:$Reqs\_Set$} 
\KwOut{Spatio-Temporal Graphs Set:$G\_Set$}
\textbf{Init} Global Graph: $GG=\emptyset$, $G\_Set=\emptyset$\\
\For{$record_i$ in $Reqs\_Set$}
{
    Add $node_i$ ($t_i$, $geo_i^7$) to $GG$ \\
    \If{$user_j$ is a friend to existing node's user $user_i$}
    {
        Add $edge_{ji}$ ($\mathcal{E}_{ji}$) to $GG$
    } 
}

\For{$node_i$ in $GG$}
{   // Avoid modeling duplicate areas \\
    \If{$geo_i^6$ hasn't appear}  
    {
        get all $nodes_j$ that $geo_j^6 == geo_i^6$\\
        // Sub-area extraction \\
        get $subgraph_i$  all $nodes_j$\\
        \textbf{Init} $G^a=\emptyset$\\
        \For{$T_i$ in Time Windows}  
        {   // Time extraction \\
            get all $nodes_j$ in $subgraph_i$ that $t_j$ in $T_i$\\
            get $\mathcal{G}^a_{T}$ contains all $nodes_j$ \\
            get $X$, $\mathcal{E}$ of $\mathcal{G}^a_{T}$ \\
            Add $\mathcal{G}^a_{T}$ to $G^a$ \\
        }
        add $G^a$ to $STG\_Set$ \\
    } 
}
\textbf{return} $STG\_Set$
\end{algorithm}


\N{\textbf{($i$) Construction of global CDSs network $\boldsymbol{GG}$.} As shown in steps 2-7, the first step is to build the global CDSs network $GG$ based on CDSs and users' social relations. Specifically, we create a node for each CDS and generate an edge between two nodes if a social connection exists between the two users (possibly friends or the CDS is a content share). Note that the edges in the CDSs network are all directed edges obtained according to the time sequence.}


\textbf{($ii$) Generation of high-dimensional features.} As shown in step 19, another goal of the algorithm is to translate the Geohash encoding of CDSs into high-dimensional features of nodes and generate edge features. We propose the node-area diffusion method to calculate node features. Specifically, when calculating the features of node $i$, we consider the impact of CDSs within its surrounding $\mathcal{P}$ areas. Therefore, the node features $X$ will be obtained as follows:
\begin{equation}
X[i][j]=\left\{
\begin{aligned}
0 & , & geo_i[\mathrm{last}] \neq j, \\
I & , & geo_i[\mathrm{last}]  =   j.
\end{aligned}
\right.
\end{equation}where $i \in [0, n$) is the node id, $j \in [0, \mathcal{P})$ is the area id, and \M{$I$ is the CDSs intensity, which is set to the number of successfully completed content requests, e.g., the number of application downloads or video transfers. Since Geohash uses 32-decimal encoding, we set $last$ to the decimal form of the last digit of the encoding, e.g., $gbsuv[last]=28$.}

The approach to generate node features based on adjacent areas has two benefits: first, the distribution of CDSs under the $\mathcal{P}$ areas can be rebuilt easily from all node features. Second, the adjacent areas that have never been recorded with CDSs can be represented and modeled.

\N{We then set the spatio distances between nodes (CDSs) as edge features $\mathcal{E}$:
\begin{align}
&\Delta lng=lng_i-lng_j \nonumber, \Delta lat=lat_i-lat_j \nonumber \\
&d=sin(\Delta lat/2)^2 + cos(lat_i) cos(lat_j) sin(\Delta lng/2)^2 \nonumber \\
&\mathcal{E}_{ij} = 2dsin(\sqrt{d})*r 
\end{align}where $lng_i$ and $lat_i$ are the longitude and latitude of node $i$, and $r=6371$km is the radius of earth.}

\N{\textbf{($iii$) Sub-area extraction and time segmentation on $\boldsymbol{GG}$.} As shown in steps 8-23, the last function of the algorithm is to perform sub-area extraction and time extraction on the global graph. The main purpose of sub-area extraction is to reduce the dimension of node features. As stated in \textbf{ii}, the feature dimension $\mathcal{P}$ of node $i$ is equal to the number of its adjacent areas. Without sub-area extraction, the node dimension that the model needs to handle would be unacceptably large. We use the encoding prefix of Geohash to extract all $\mathcal{P}$ adjacent areas (called an \textbf{unit}) of an area (as shown in steps 11-13). Each new bit encoded by Geohash will divide the previous area into 32 sub-areas, so we set $\mathcal{P}=32$. The purpose of time extraction is to distinguish CDSs networks under different time windows, which is common in time-series prediction problems.}


\begin{figure}[tbp]
    \centering
    \includegraphics[width=\columnwidth]{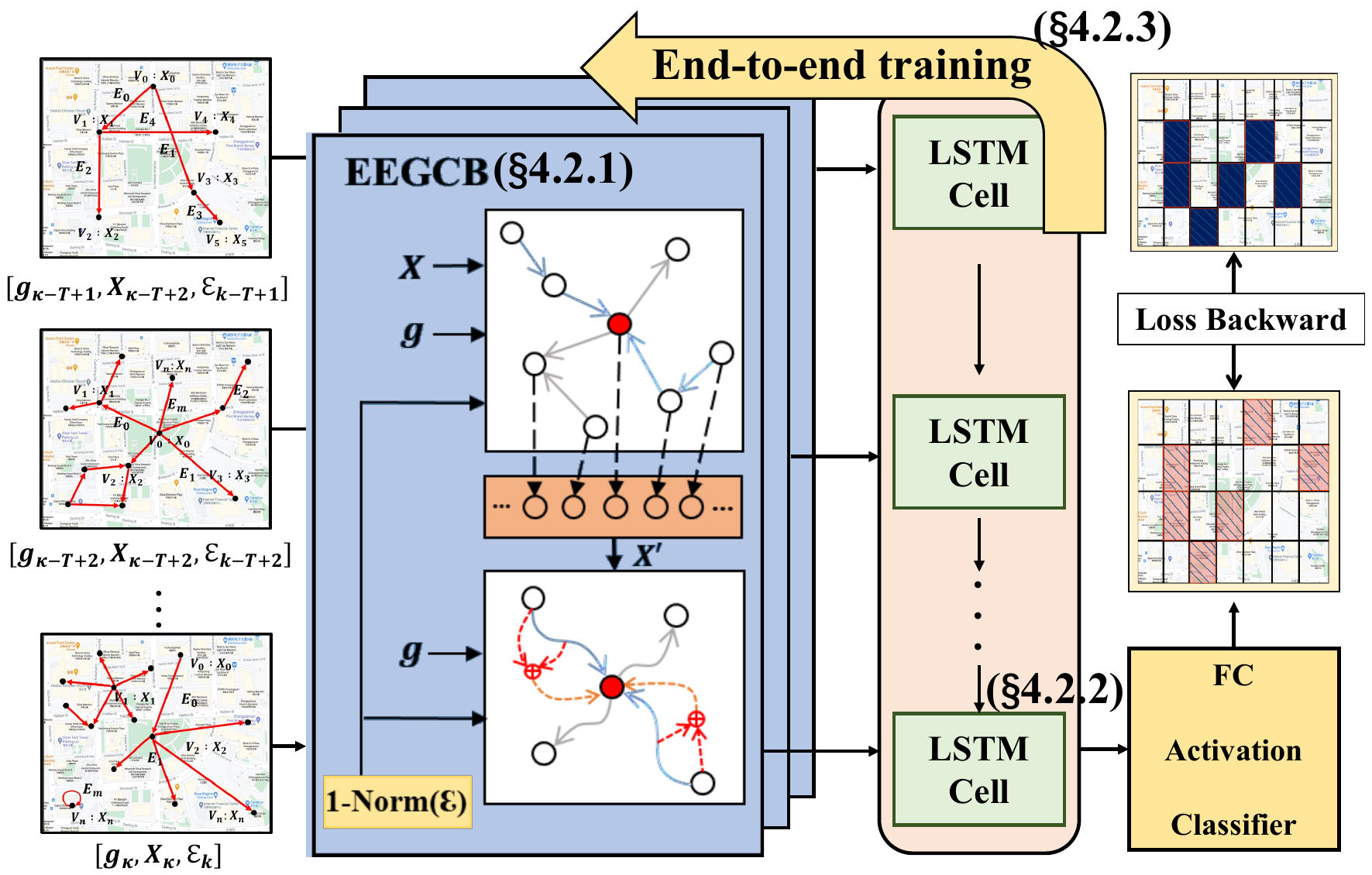}
    \caption{T-EEGCN Framework}
    \label{fig:teegcn}
\end{figure}

\subsection{Multi-feature Based Prediction Framework}
\N{As Fig. \ref{fig:teegcn} shows, the proposed Temporal-Edge Enhanced Graph Convolution Network (T-EEGCN) consists of four components: input and edge attribute normalization, Edge-Enhanced Graph Convolution Block (EEGCB) for spatio-social features extraction, Long Short Time Memory (LSTM) for temporal dependency extraction, and fully connected layer for result formalization. In the flowing parts, we will explain how the T-EEGCN is adopted to realize the spatio fine-grained hotspots prediction task.}


\begin{figure}[b]
    \centering
    \includegraphics[width=0.9\columnwidth]{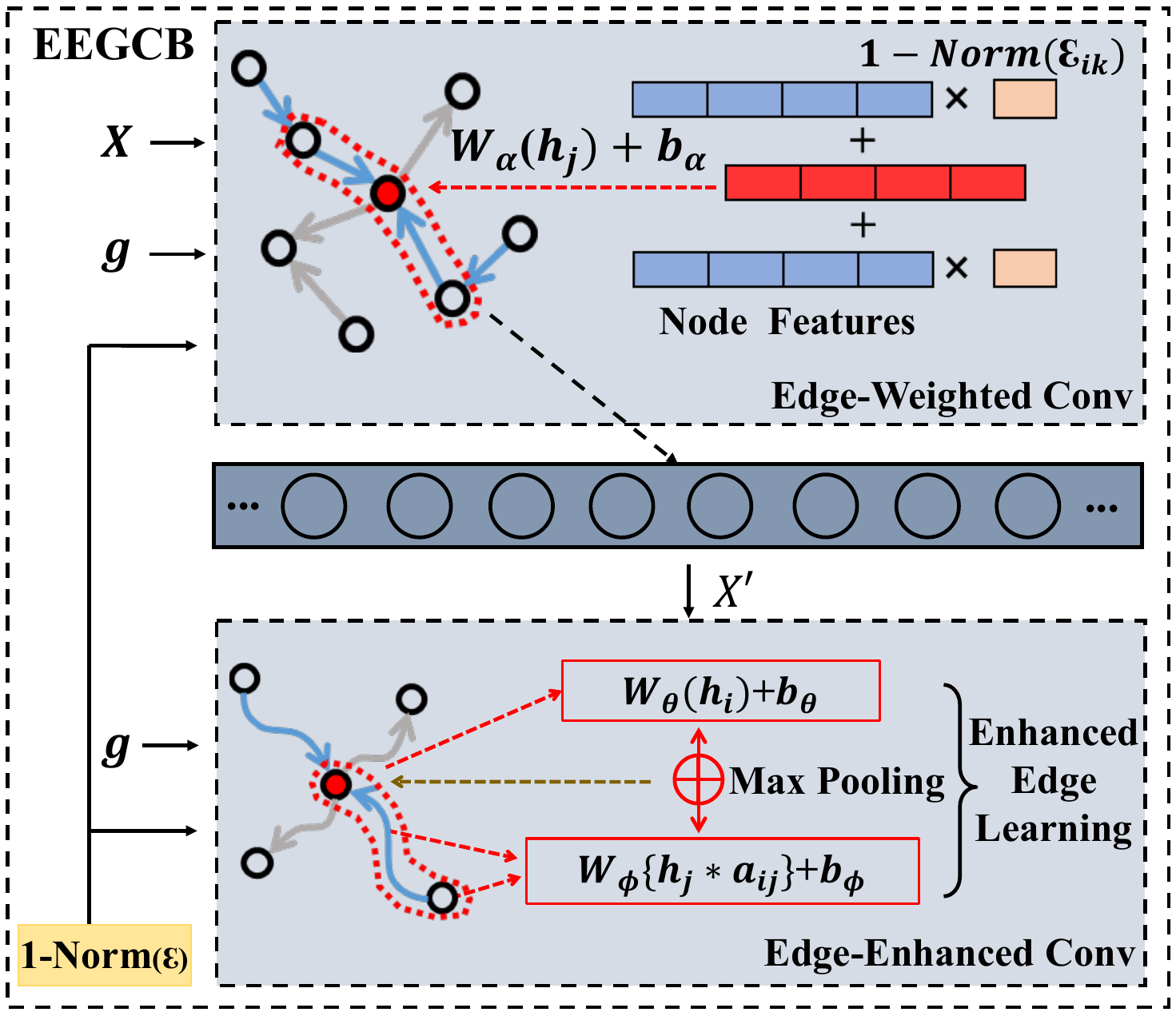}
    \caption{Edge-Enhanced Graph Convolution Block}
    \label{fig:eegcb}
\end{figure}

\subsubsection{Edge-Enhanced Graph Convolution Block for Extracting Spatio-Social Features}
As shown in Fig. \ref{fig:eegcb}, we design two graph convolution layers in EEGCB and introduce three types of learnable parameters (\textbf{with bold font}) to adjust the propagation and aggregation functions of the graph convolution, enabling EEGCB to handle dynamic graphs flexibly. The first edge-weighted convolution layer is used to update node embedding based on social relations and spatio distances. The second edge-enhanced convolution layer will further enhance the EEGCB's capability to learn the spatio features and handle dynamic graphs. 


\textbf{Edge-Weighted Graph Convolution.}
\label{model:ew}
The topology structure of the CDSs network represents the \textbf{social relations} between users who led to these CDSs. GCN can naturally learn this structure feature through the message passing and aggregation of neighbors' information of the central node \cite{kipf2017semisupervised}. However, in our problem, the node embedding vectors $H=\{h_0,\ldots,h_n\} \in \mathbb{R}^{n \times \mathcal{P}'}$ in convolution ($n$ is the number of nodes and $P'$ is the feature dimension) represents the spatio distribution of CDSs. Therefore, we need to learn how the social relations will affect the associated CDSs' spatio distribution, i.e. the weight of node features influenced by neighbors' features.

Since the CDSs networks are embedded in real geographic space, a deserved way is to add spatio distance as edge feature $A \in \mathbb{R}^{n \times n}$. We design an edge-weighted graph convolution layer to use the \textbf{spatio distance} to adjust the influence of neighboring nodes $\mathcal{N}(i)$ on the features of the central node $i$. The layer-wise propagation rule can be defined as follows:

\begin{equation}
\label{ew}
    h_i^{l+1} = \sigma(b^l + \underset{j\in \mathcal{N}(i)}{\sum}a_{ij}h_j^{l}W^l_{\alpha})
\end{equation}where \M{$l$ represents the layer of the convolution}, $h_i$ is the feature of node $i$, $a_{ij}$ is the edge weight between node i and j, $\boldsymbol{W_{\alpha}}$ is the first learnable parameter used to adjust the information aggregation rules based on the dynamic graph structures while the propagation rules are determined (i.e., weights on edges), and $\sigma$ is an activation function.

Referring to the subsequent data analysis and study \cite{scellato2011socio}, we know that the social closeness will decrease with the spatio distance increase. To this end, we set the edge weight $a_{ij}$ to the negative normalized Euclidean distance with bias:

\begin{equation}
    a_{ij} = 1-\mathrm{Norm}(\mathcal{E}_{ij}) = 1-\frac{\mathcal{E}_{ij} - min(\mathcal{E}_{i:})}{max(\mathcal{E}_{i:})-min(\mathcal{E}_{i:})}
\end{equation}where $\mathcal{E}$ is the input edge attribute.

Users' requests are affected by friends and the entire group. The influence will decrease with the increase of distance. The edge-weighted graph convolution will capture the two types of dependency for each node through information propagation and aggregation.



\textbf{Edge-Enhanced Graph Convolution.}\label{model:eegc} Although exploiting the property that social closeness decreases as distance increases, the intuitive weights cannot wholly represent the relationship between CDSs. Due to user mobility and online friendships, sometimes there are \emph{long-distance} spatio dependency between CDSs. Besides, with the structures and features of the \textbf{dynamic graphs} change, both information propagation and aggregation rules need to be adjusted to enhance the learning of the dynamic graphs. 

Therefore, we design an edge-enhanced graph convolution layer to enhance the learning of \textbf{spatio features} and the flexibility of graph convolution. The edge-enhanced graph convolution layer also accepts external inputs as edge features. Here we input the normalized distance $\mathrm{Norm}(\mathcal{E})$, but do not directly treat it as the edge weight. We introduce the second learnable parameter $\boldsymbol{W_{\phi}}$ to dynamically adjust the influence of neighboring node $j$ at different distances on the node being updated (adjust the information propagation rules): 

\begin{equation}
    m_{ij}^l = W_{\phi}(h_j^l * \mathrm{Norm}(\mathcal{E}_{ij}))+b_{\phi}
\end{equation}where $m_{ij}$ represents the update information to node $i$ from node $j$, $b_{\phi}$ are the bias of the linear layer, in this way, $m$ is not only influenced by the distance between nodes but also varies with the features of neighboring nodes.

Since the central node i has already encoded the neighbors' information from the first convolution layer, we introduce the third learnable parameter $\boldsymbol{W_{\theta}}$ to preserve the learned node features $h_i$ dynamically (adjust the information aggregation rules):

\begin{equation}
    h_{i}'^l = W_{\theta}(h_i^l)+b_{\theta}
\end{equation}where $b_{\theta}$ are the bias. Finally, the propagation rule for the edge-enhanced graph convolution layer can be formalized as follows:

\begin{equation}
\label{ee}
    h_i^{l+1} = \max_{j \in \mathcal{N}(i)} \mathrm{ReLU}(m_{ij}^l+h_{i}'^l)
\end{equation}where $\max$ is the max pooling. At the end of T-EEGCN, to obtain the embedding representation $\mathcal{V}$ of the graph, we average the feature vectors of all nodes as the output, i.e. $\mathcal{V}_t = \sum_{i \in V_t}h_i/|V_t|$. 

\subsubsection{Long Short Term Memory for Extracting Temporal Features}
The occurrence of CDSs is often affected by users' historical behavior, e.g., watching regularly updated videos and downloading games during holidays. This temporal dependency leads to the correlations of CDSs networks under different time windows, and GCN cannot capture this distinction and connection.

Currently, the most widely used neural network model for temporal dependency extraction is the Recurrent Neural Network (RNN). However, due to defects such as gradient disappearance and explosion, the traditional RNN has limitations for long-term prediction. As a variant of RNN, Long Short Term Memory (LSTM) \cite{10.1162/neco_a_01199} has been proven to solve the above problems through the gates function.

Here, we adopt LSTM to capture the \textbf{temporal dependency} between CDSs networks at different time windows. The input of LSTM is a sequential graph embedding vector $\boldsymbol{\mathcal{V}} = \{\mathcal{V}_{\kappa-T+1},\ldots,\mathcal{V}_{\kappa}\}$, the compute process for each layer can be described as follows:

\begin{align}
\label{eq:LSTM}
    &f_t = \sigma(W_f\cdot[h_{t-1},x_t]+b_f) \nonumber\\
    &i_t = \sigma(W_i\cdot[h_{t-1},x_t]+b_i) \nonumber\\
    &C_t = f_t \ast C_{t-1}+i_t \ast tanh(W_C\cdot[h_{t-1},x_t]+b_C) \nonumber\\
    &o_t = \sigma(W_o\cdot[h_{t-1},x_t]+b_o) \nonumber\\
    &h_t = o_t \ast tanh(C_t)
\end{align}where $t$ is the time step in terms of the length of time windows, $h_t,c_t,x_t$ are the hidden state, cell state, and input at time $t$ ($x_t = \mathcal{V}_{g_t}$), and $f_t,i_t,o_t$ are respectively the forget gate, input gate and output gate.

\subsubsection{Calculation Process and Result Formalization}
The calculation process of T-EEGCN can be seen in Fig. \ref{fig:teegcn}. The graph topology, node features, and edge features will be input into the Edge-Enhanced Graph Convolution Block to produce a sequence of graph representation vectors $\boldsymbol{\mathcal{V}}$. After that, $\boldsymbol{\mathcal{V}}$ will be input into the LSTM for temporal dependency extraction (equation.\ref{eq:lstm}), and finally the output of LSTM will be passed to FC layer and sigmoid function (equation.\ref{eq:result}) to generate the hotspots matrix $\hat{M}$, which can be formalized as follows:


\begin{align}
    &X'_t = \sigma(AX_t W_{\alpha})\\
    &\mathcal{V}_t = mean(\sigma(W_{\theta}X'_tW_{\phi}))\\
    &\boldsymbol{\mathcal{V}}=\{\mathcal{V}_{\kappa-T+1},...,\mathcal{V}_{\kappa}\}\\
    \label{eq:lstm}
    &O = \mathcal{L}(\boldsymbol{\mathcal{V}}, W_L)\\
    \label{eq:result}
    &\hat{M} = \mathrm{sigmoid}(W O_i+b)
\end{align}where $A$ is the edge weight matrix, $W_L$ represents the trainable parameters of the LSTM, $W$ and $b$ represent the parameter matrix and bias of FC, respectively. We use the BCELoss to measure the performance of our model which is defined as follows,

\begin{equation}
    L(\hat{M}, M) = -W[M log(\hat{M}) + (1-M)log(1-\hat{M})]
\end{equation}where $W$ is the weight matrix. 

We summarize the main characteristics of T-EEGCN in the following,
\begin{itemize}
    \item T-EEGCN is a universal framework for processing dynamic graphs. It can not only tackle CDSs networks modeling and prediction issues but also be applied to more general spatio-temporal sequence learning tasks.
    \item We introduce three learnable parameters in EEGCB and guide the update of the parameters based on the dynamic graph structures and edge features. The design allows EEGCB to adjust the node embedding update rules based on social and spatio features. EEGCB does not depend on the structure information of a specific graph and can be applied to dynamic CDSs networks coding.
    \item The output layer of EEGCB is tightly connected to the input layer of LSTM. While the hybrid model is being trained, the parameters of EEGCB and LSTM can be jointly updated. Thus, the whole framework can be trained end-to-end.
\end{itemize}


\subsubsection{Computational Complexity}

\N{In the analysis of computational complexity, we assume that the number of hidden units in EEGCB is constant. The size of the time window $T$ is a fixed number during the training and evaluation process. These parameters have no relationship with the input scale, so we don't consider them. Then the computational complexity of T-EEGCB can be represented as $\mathcal{O}(\lvert V \rvert \cdot \lvert E \rvert \cdot (\mathcal{P} + \lvert \mathcal{E} \rvert))$, where:}

\N{\begin{itemize}
    \item $\lvert V \rvert$ represents the number of nodes in every CDSs network $\mathcal{G}_j^i$. Since we want to aggregate the nodes in a CDSs network, we should traverse every node in $\mathcal{G}_j^i$.
    \item $\lvert E \rvert$ represent the number of edges in every CDSs network $\mathcal{G}_j^i$. For every node in $V$, we use node sampling to get the node sets. In this study, we sample the $K = 2$ depth of neighbor nodes. However, the $K = 2$ neighbor nodes are uncertain but there exists an upper bound $\lvert E \rvert$ of the neighbor nodes, in which $\lvert E \rvert$ represents the number of edges.
    \item $\mathcal{P} + \lvert \mathcal{E} \rvert$ represents the cost of aggregation and edge enhance, in which $\mathcal{P}$ represents the dimension of node features and $\lvert \mathcal{E} \rvert$ represents the dimension of edge features. Since aggregating is applied to node features and edge enhance is applied to edge features, and only simple operations in the aggregation and edge enhance the process, the complexity of the two processes can be denoted $\mathcal{O}(\mathcal{P})$ and $\mathcal{O}(\lvert \mathcal{E} \rvert)$ respectively.
\end{itemize}}

\section{Experiments}\label{Exp}
\subsection{Dataset Description}
We use two large-scale real-world mobile content datasets: point-to-point (P2P) file transmission records \textbf{OPPST} and online places check-in sharing data \textbf{Gowalla}\footnote{http://snap.stanford.edu/data/loc-Gowalla.html} in this paper.

\textbf{OPPST} comes from the log files of a P2P file sharing app, whose transferred contents include not only files and pictures but also videos and apps. The trace is 843GB, including 30485335 users and 443440043 sharing records, ranging from 08/01/2016 to 10/30/2016.

\textbf{Gowalla} is a location-based social network created in 2009. The users check in at places through their mobile devices. Check-ins are shared with friends, and the dataset contains a total of 6,442,890 check-ins of users over the period of Feb. 2009 - Oct. 2010.

\subsection{Data Analysis}
\label{data_pro}

\begin{figure}
\centering
\subfigure[Spatio Correlation]{\label{fig:spatio}
\includegraphics[height=3.1cm]{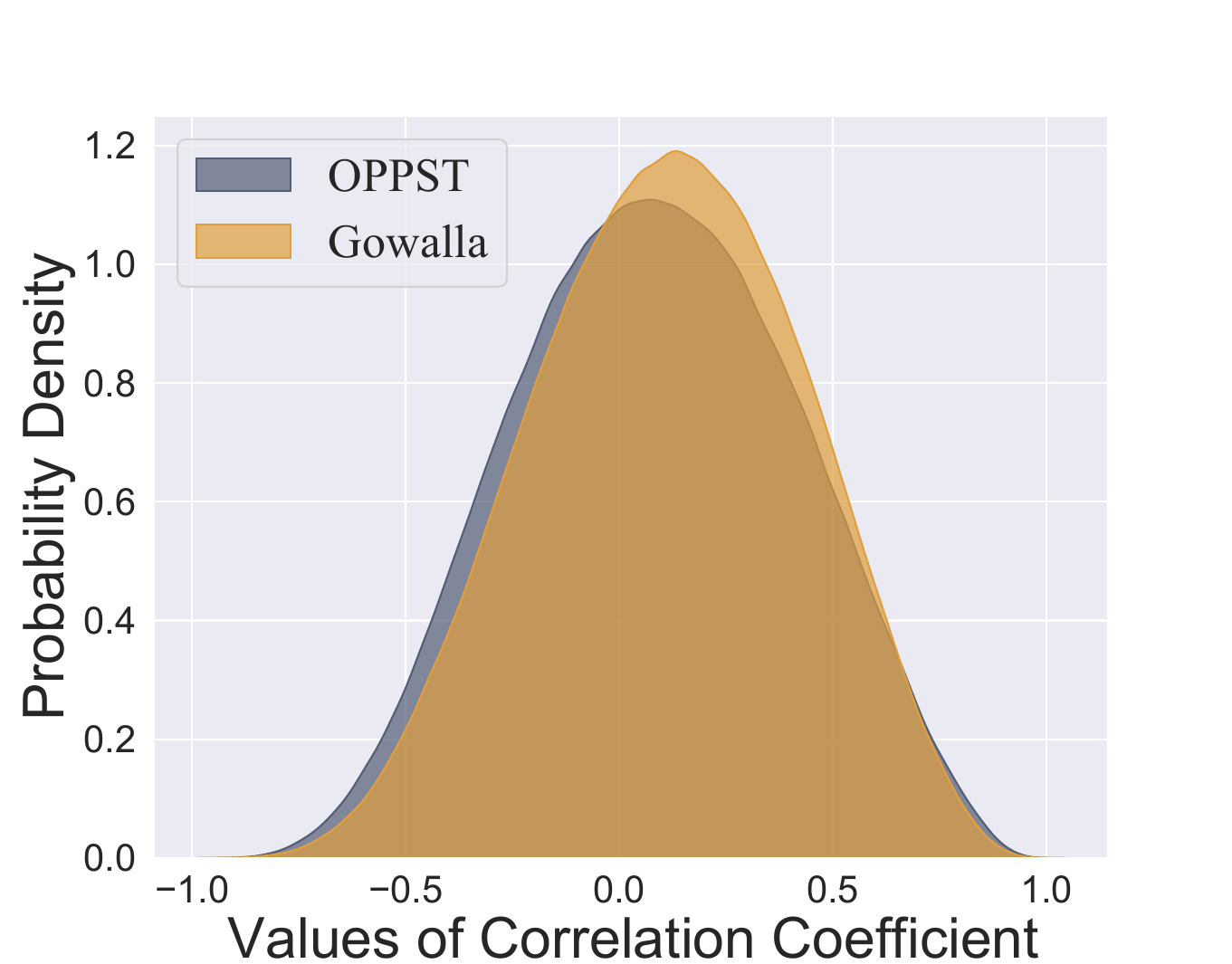}
}
\subfigure[CDF of Social Dis.]{\label{fig:social}
\includegraphics[height=3.1cm]{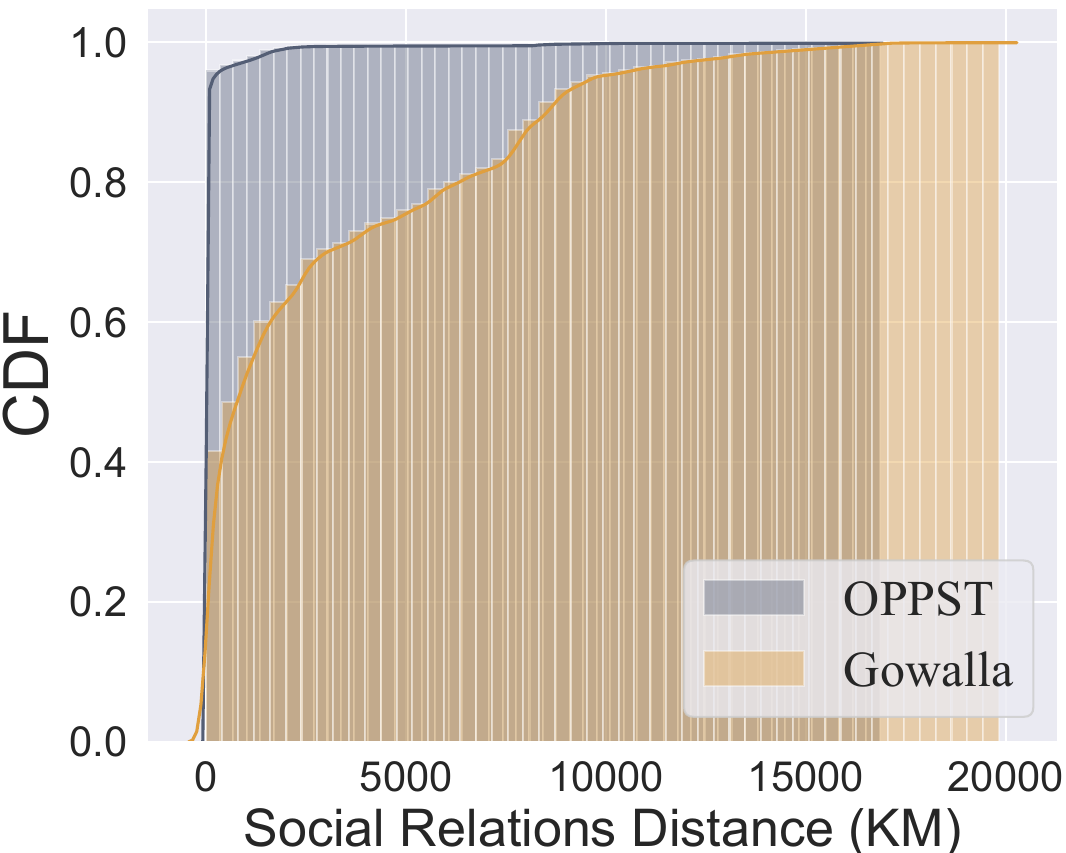}
}
\subfigure[Fluctuation of CDSs Over Time]{\label{fig:time}
\includegraphics[height=3.4cm]{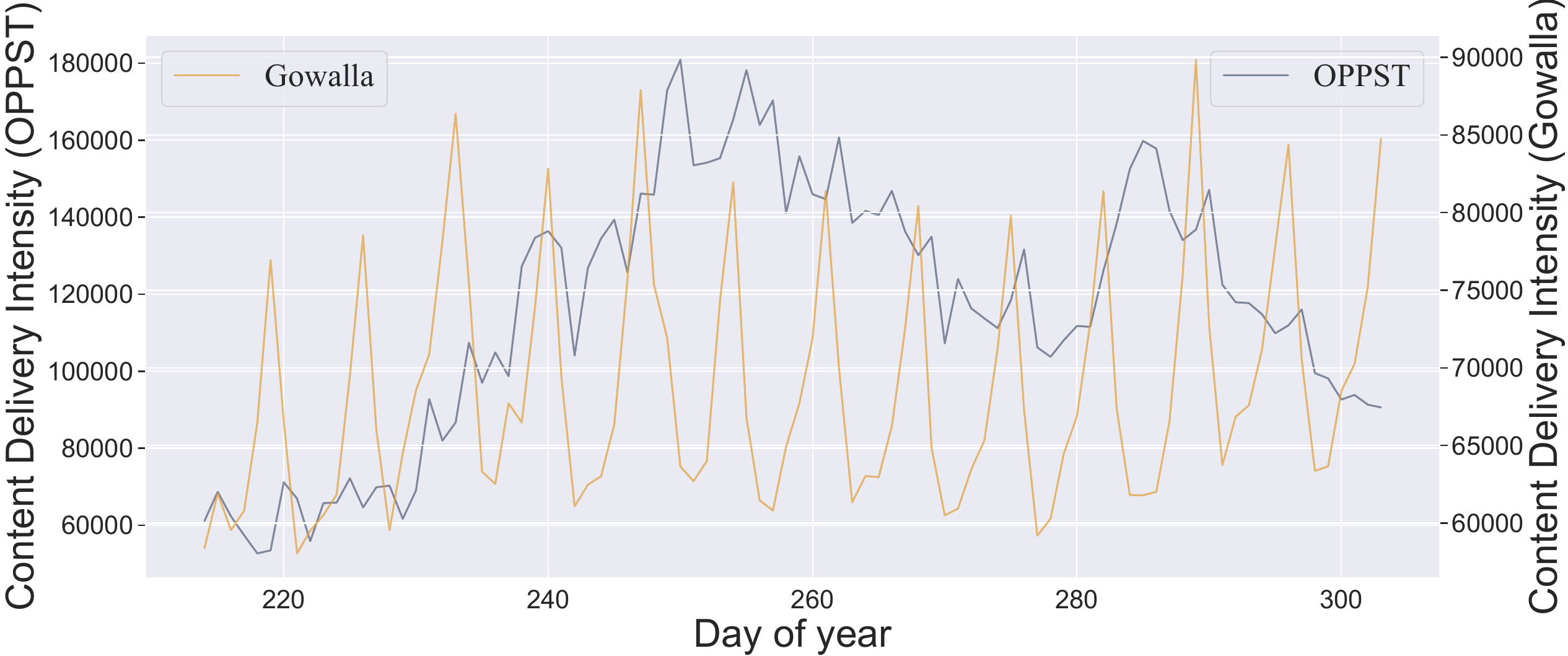}
}
\caption{Spatio-Temporal-Social Dependencies Analysis}
\label{fig:feature}
\end{figure}

As shown in Fig. \ref{fig:feature}, we examined the spatio correlation, social distance distribution, and temporal dependency of CDSs in both datasets. Our main observations are summarized as follows.

\N{\textbf{Observation 1: The distribution of CDSs are spatio correlated.}} Fig. \ref{fig:spatio} shows the probability distribution of the Pearson correlation coefficient of CDSs between different areas (within 76m). It can be seen that the correlation coefficients of the OPPST are mainly concentrated in 0.1, showing a certain non-zero correlation in the spatio domain. On the other hand, the correlation coefficients of Gowalla are higher, indicating that its spatio features have a more substantial influence on CDSs.

\N{\textbf{Observation 2: The distribution of CDSs is intensely temporal dependent.}} Fig. \ref{fig:time} shows that the CDSs of both datasets have a solid seasonal periodic feature within a year, especially Gowalla, which has a significant cyclical variation, which proves that the distribution of CDSs has a solid temporal dependency.

\N{\textbf{Observation 3: Social dependency decreases with distance.}} It can be seen from Fig. \ref{fig:social} that the social distances of users among most CDSs are short ($<$1000km), which is consistent with our setting in Section \ref{model:eegc}. However, there are also long-distance social relationships that exceed 10000km, so dynamic adjustment of the edge features is needed. Since OPPST is a D2D-type transmission dataset, its social distance is much smaller than that of the Gowalla, which is shared online.

The above analysis demonstrates the spatio-temporal-social dependencies of CDSs and validates the T-EEGCN design's rationality from the perspective of data regularity.

\subsection{Experimental Settings}
We set the number of periods for training to 9 ($T=9$) and the number for prediction to 1. Thus, the length of each time window is seven days in OPPST and thirty days in Gowalla.

\textbf{Chosen Categories.} Different content categories of CDSs significantly affect their spatio-temporal distribution patterns \cite{article}. To better focus on the effects of spatio-temporal-social dependency, we select four major categories of app files under the OPPST and three categories of most frequent check-in places under the Gowalla for experiments.

\N{\textbf{Networks and Spatio Information.} Table \ref{data infor} illustrates the total number of nodes and edges of the spatio-temporal graphs utilized in the experiment, which are sufficient to support the training of end-to-end models. Table \ref{spatio information} illustrates the number of units (refer to $\S$4.1 \#iii), the total number of areas, the number of known areas, and the percentage of hotspots in the time window to be predicted, respectively.}



\begin{table}
\caption{Networks Information}
\label{data infor}
\resizebox{8.5cm}{!}{
\footnotesize
\centering
\begin{tabular}{llcc}
\hline
\textbf{DataSets} &\textbf{Cate.} &\textbf{Nodes} &\textbf{Edges}  \\ \hline 
OPPST   &Game(OPP1)     &1,791,310 &2,133,529    \\
        &Tools(OPP2)    &1,072,962 &1,306,140    \\
        &Video(OPP3)    &1,521,456 &1,746,311    \\
        &Comm.(OPP4)    &1,305,664 &1,472,994    \\  \hline
Gowalla &Coffee(Gow1)   &3,999,797 &10,614,556   \\
        &Sand.(Gow2)    &1,927,685 &4,341,135    \\  
        &Theatre(Gow3)  &917,141  &1,393,035    \\   \hline
\end{tabular}}
\end{table}


\begin{table}
\caption{Spatio Information}
\label{spatio information}
\resizebox{8.5cm}{!}{
\footnotesize
\centering
\begin{tabular}{ccccc}
\hline
\textbf{Cate.} &\textbf{Units} &\textbf{Areas}  &\textbf{Known Areas} &\textbf{Hot Rate($\tau$)}  \\ \hline 
OPP1     &5,756  &184,224  &39,279 &9.1\%(3)   \\
OPP2     &1,926  &61,632   &9,864  &8.5\%(5)   \\
OPP3     &4,255  &136,160  &26,748 &10.4\%(5)  \\
OPP4     &2,205  &70,560   &13,974 &10.0\%(3)  \\  \hline
Gow1     &10,285 &329,120  &46,486 &14.6\%(3)  \\
Gow2     &5,519  &176,608  &24,939 &9.1\%(3)  \\  
Gow3     &12,149 &388,768  &56,188 &6.6\%(5)  \\   \hline
\end{tabular}}
\end{table}

\begin{table*}[h]
\caption{Performance Comparison of Different Approaches on OPPST and Gowalla.}
\label{result}
\resizebox{\textwidth}{!}{
\begin{tabular}{c|cccc}
\hline
\multirow{2}{*}{Model} & \multicolumn{4}{c}{OPPST(OPP1/OPP2/OPP3/OPP4)}                                                                        \\ \cline{2-5} 
                       & Prec                        & Recall                      & F1                          & AUC                         \\ \hline
SVM                    & 0.35 / 0.31 / 0.31 / 0.36 & 0.40 / 0.35 / 0.40 / 0.39 & 0.37 / 0.33 / 0.35 / 0.37 & 0.63 / 0.61 / 0.61 / 0.63 \\
DeepInf                & 0.42 / 0.33 / 0.44 / 0.38 & 0.30 / 0.23 / 0.24 / 0.19 & 0.35 / 0.27 / 0.31 / 0.26 & 0.63 / 0.59 / 0.59 / 0.58 \\
D2D-LSTM               & \underline{0.59} / \underline{0.61} / 0.56 / \underline{0.61} & \underline{0.52} / \underline{0.55} / \underline{0.48} / \underline{0.55} & \underline{0.56} / \underline{0.58} / \underline{0.52} / \underline{0.58} & \underline{0.74} / \underline{0.76} / \underline{0.72} / \underline{0.75} \\
T-GCN               & 0.51 / 0.55 / 0.55 / 0.51 & 0.37 / 0.32 / 0.28 / 0.43 & 0.43 / 0.40 / 0.37 / 0.47 & 0.63 / 0.62 / 0.62 / 0.64 \\
MT-GCN               & 0.54 / 0.59 / \underline{0.57} / 0.56 & 0.39 / 0.35 / 0.29 / 0.43 & 0.46 / 0.44 / 0.38 / 0.49 & 0.63 / 0.62 / 0.63 / 0.65 \\
\textbf{T-EEGCN}                & \textbf{0.76} / \textbf{0.76} / \textbf{0.76} / \textbf{0.76} & \textbf{0.64} / \textbf{0.63} / \textbf{0.64} / \textbf{0.65} & \textbf{0.70} / \textbf{0.68} / \textbf{0.69} / \textbf{0.70} & \textbf{0.84} / \textbf{0.85} / \textbf{0.84} / \textbf{0.84}  \\ \hline
\multirow{2}{*}{Model} & \multicolumn{4}{c}{Gowalla(Gow1/Gow2/Gow3)}                                                                               \\ \cline{2-5} 
                       & Prec                        & Recall                      & F1                          & AUC                         \\ \hline 
SVM                    & 0.75 / 0.76 / 0.77        & 0.46 / 0.46 / 0.50        & 0.57 / 0.58 / 0.61        & 0.62 / 0.62 / 0.65        \\
DeepInf                & 0.87 / 0.78 / 0.82        & 0.72 / 0.70 / 0.65        & 0.79 / 0.74 / 0.72        & 0.85 / 0.84 / 0.82         \\
D2D-LSTM               & \textbf{0.95} / 0.87 / 0.84        & \underline{0.80} / \underline{0.74} / \underline{0.69}        & \underline{0.87} / \underline{0.80} / \underline{0.75}        & \textbf{0.91} / \underline{0.86} / \underline{0.85}        \\
T-GCN               & 0.81 / 0.874 / 0.88        & 0.43 / 0.43 / 0.53        & 0.56 / 0.58 / 0.66        & 0.75 / 0.71 / 0.76        \\
MT-GCN               & 0.82 / \underline{0.88} / \underline{0.89}        & 0.44 / 0.44 / 0.54        & 0.57 / 0.59 / 0.67        & 0.76 / 0.72 / 0.77        \\
\textbf{T-EEGCN}                & \underline{0.94} / \textbf{0.88} / \textbf{0.90}        & \textbf{0.82} / \textbf{0.80} / \textbf{0.82}        & \textbf{0.87} / \textbf{0.84} / \textbf{0.86}        & \underline{0.91} / \textbf{0.91} / \textbf{0.91}         \\ \hline
\end{tabular}}
\end{table*}

\textbf{Evaluation Metric \& Baselines.} We adopted Precision (Prec), Recall, F1, and Area Under the ROC Curve (AUC) to evaluate the performance. The baselines are as follows:

\begin{itemize}
    \item SVM \cite{platt1999probabilistic}: Support Vector Machine method, widely used in classification tasks.
    \item DeepInf \cite{qiu2018deepinf} A GCN/GAT-based framework for learning social relations of users' social networks. Here we use its GCN form to capture the social features of the CDSs networks.
    \item D2D-LSTM \cite{article} A LSTM-based recurrent network takes multiple dimensions into account, including time, geography, and category.
    \item T-GCN \cite{Zhao2020} A temporal GCN for static road network traffic prediction, which is combined with the GCN and the gated recurrent unit (GRU).
    \item Macro Temporal GCN (MT-GCN) \cite{liu2019characterizing} A temporal GCN for social influence prediction, which can encode the dynamic social networks and capture the temporal feature.
\end{itemize}

\subsection{Evaluation}\label{er}

\textbf{Fine-grained hotspots prediction accuracy.} As shown in Table \ref{result} and Table \ref{result_avg} \footnote{The bold in Tables 5 and 6 means the optimum value in a column, and underlining represents the suboptimal value, and we compare the 0.001 level when the two values are equal.}, T-EEGCN performs well in fine-grained hotspots prediction for four evaluation metrics. Specifically, it performs better than the five baselines on the different categories and the average values of the two datasets. Numerically, it significantly outperforms other models at the 0.01 level. T-EEGCN performs very well on Gowalla for \emph{long-term} prediction, achieving an average F1 of up to 0.86 and an average AUC of 0.91. 

Unlike T-EEGCN, DeepInf cannot handle dynamic graphs. It deals with the network of all CDSs in the first nine periods and computes the graph structure feature once to update the final node embedding. Besides, DeepInf does not exploit edge features to adjust the update rules. The drawbacks of these two mechanisms make DeepInf inappropriate for extracting spatio-temporal features, leading to its poor performance (0.49/0.24 for average F1/AUC in OPPST and 0.11/0.07 for average F1/AUC in Gowalla compared to T-EEGCN). It indicates that the single social feature has an insufficient influence on predicting CDSs hotspots. The real-world spatio-temporal features need to be considered. 

D2D-LSTM achieves sub-optimal performance in most content categories, which indicates that CDSs have a strong temporal dependency, i.e., users are more likely to request mobile content in areas where they have historically performed requests. T-EEGCN does not outperform D2D-LSTM in Prec and AUC in Gow1. It is because Gow1 is the LBSN of category \textbf{Coffee}. As analyzed in Fig. \ref{fig:time}, CDSs under Gowalla has a strong temporal periodicity, so the model's ability to extract temporal features plays a greater role (at which D2D-LSTM is good). However, T-EEGCN outperforms D2D-LSTM in all other datasets, especially in Gow3 \textbf{(Theatre)}, T-EEGCN exceeded D2D-LSTM by 0.06, 0.13, 0.11, and 0.06 for the four metrics. It is because CDSs about Theatre are less affected by temporal features compared to daily communication behaviors, e.g., checking in at \textbf{Coffee} (Gow1) and \textbf{Sandwich shop} (Gow2).

Compared with D2D-LSTM, T-EEGCN can capture spatio-social features through the encoding of CDSs networks by EEGCB. Although D2D-LSTM introduces the influence of spatio features by embedding geographic location, it does not consider the \textbf{social relations} between users, making its prediction ability weaker than T-EEGCN (0.14/0.10 for average F1/AUC in OPPST and 0.05/0.03 for average F1/AUC in Gowalla).

The metrics show that MT-GCN outperforms T-GCN, it is because the GCN of T-GCN is applied to the static graph, which computes the graph structure once and updates the node embedding according to the fixed graph structure. However, the GCN of MT-GCN is trained to handle dynamic social networks. It adjusts node embedding rules according to different graph structures—the above results prove the necessity for dynamic graph learning. 

Although both MT-GCN and T-EEGCN are combinations of GCNs and recurrent neural networks, compared to MT-GCN, T-EEGCN has significantly improved the prediction performance on all datasets (0.25/0.21 for average F1/AUC in OPPST and 0.25/0.16 for average F1/AUC in Gowalla). Compared to the GCN of MT-GCN, the EEGCB of T-EEGCN can update node embedding by utilizing both dynamic graph structures and edge features, enabling EEGCB to capture fine-grained spatio dependency features, thus encoding CDSs networks more accurately.

\N{\textbf{Generalizability analysis for unknown areas prediction.} From Table \ref{spatio information} we can calculate the percentage of known areas in different datasets, where the average percentage of known areas in OPPST accounts for 19\%, while Gowalla is 14\%. In addition, due to the truncation of the hotspots threshold $\tau$, the Hot Rate under the last time window will be smaller than the percentage of known areas (9.5\% for OPPST and 10.1\% for Gowalla).}

\N{According to our setting $\mathcal{P}=32$, the percentage of areas with CDSs (known areas) should be about 3.13\%. However, the actual percentage of known areas and the Hot Rate for each dataset are much larger than 3.13\%. This phenomenon verifies that part of the areas adjacent to known areas is showing up with CDSs, even up to the size of the number of hotspots threshold.}

\N{We argue that this phenomenon occurs mainly due to the migration of CDSs caused by user mobility, as stated in \emph{challenge ii}. The large number of unknown areas poses a challenge to the spatio generalization of the model.}

\N{By combining the performance of OPP1-OPP4/Gow1-Gow3 and the corresponding percentage of unknown areas, we find that the prediction of unknown areas increases the problem's difficulty. Specifically, the higher the percentage of unknown areas in the dataset, the lower the prediction accuracy of the models tends to be. For example, OPP2 has 84\% of unknown areas, the highest percentage among the four data categories, and correspondingly all models perform worse on this dataset (except D2D-LSTM and our T-EEGCN), since D2D-LSTM is the only model that does not consider user CDSs networks, unknown areas due to user mobility have no effect on its performance.}


\N{Table \ref{result} and Table \ref{result_avg} show that for OPPST, T-EEGCN successfully predicts 76\% of the hotspots and recalls 65\% of the hotspots despite the high percentage of unknown areas of 80\%. As for Gowalla (unknown areas of 86\%), TEEGCN successfully predicts 91\% of the hotspots and recalls 81\% of the hotspots, demonstrating the strong generalization of T-EEGCN toward unknown areas. This is due to the proposed EEGCB, which learns user spatio location features, while the incorporated edge weights capture the impact of user mobility on CDSs by adapting the node update process through the spatio distance between CDSs.}

\begin{table}[t!]
\centering
\caption{Average Performance Comparison}
\label{result_avg}
\resizebox{!}{2.5cm}{
\begin{tabular}{c|llll}
\hline
\multirow{2}{*}{Model} & \multicolumn{4}{c}{OPPST\_AVG.}                                                                          \\ \cline{2-5} 
                       & \multicolumn{1}{c}{Prec} & \multicolumn{1}{c}{Recall} & \multicolumn{1}{c}{F1} & \multicolumn{1}{c}{AUC} \\ \hline
DeepInf                & 0.39                   & 0.24                     & 0.30                 & 0.60                  \\
D2D-LSTM               & 0.59                   & 0.53                     & 0.56                 & 0.74                  \\
T-GCN                  & 0.53                   & 0.35                     & 0.42                 & 0.63                  \\
MT-GCN                 & 0.56                   & 0.37                     & 0.44                 & 0.63                  \\
\textbf{T-EEGCN}       & \textbf{0.76}          & \textbf{0.64}            & \textbf{0.69}        & \textbf{0.84}         \\ \hline
\multirow{2}{*}{Model} & \multicolumn{4}{c}{Gowalla\_AVG.}                                                                        \\ \cline{2-5} 
                       & \multicolumn{1}{c}{Prec} & \multicolumn{1}{c}{Recall} & \multicolumn{1}{c}{F1} & \multicolumn{1}{c}{AUC} \\ \hline
DeepInf                & 0.82                   & 0.69                     & 0.75                 & 0.84                  \\
D2D-LSTM               & 0.89                   & 0.74                     & 0.81                 & 0.88                  \\
T-GCN                  & 0.85                   & 0.46                     & 0.60                 & 0.74                  \\
MT-GCN                 & 0.86                   & 0.47                     & 0.61                 & 0.75                  \\
\textbf{T-EEGCN}       & \textbf{0.91}          & \textbf{0.81}            & \textbf{0.86}        & \textbf{0.91}         \\ \hline
\end{tabular}}
\end{table}

\begin{figure}[h]
    \centering
    \includegraphics[width=8.7cm]{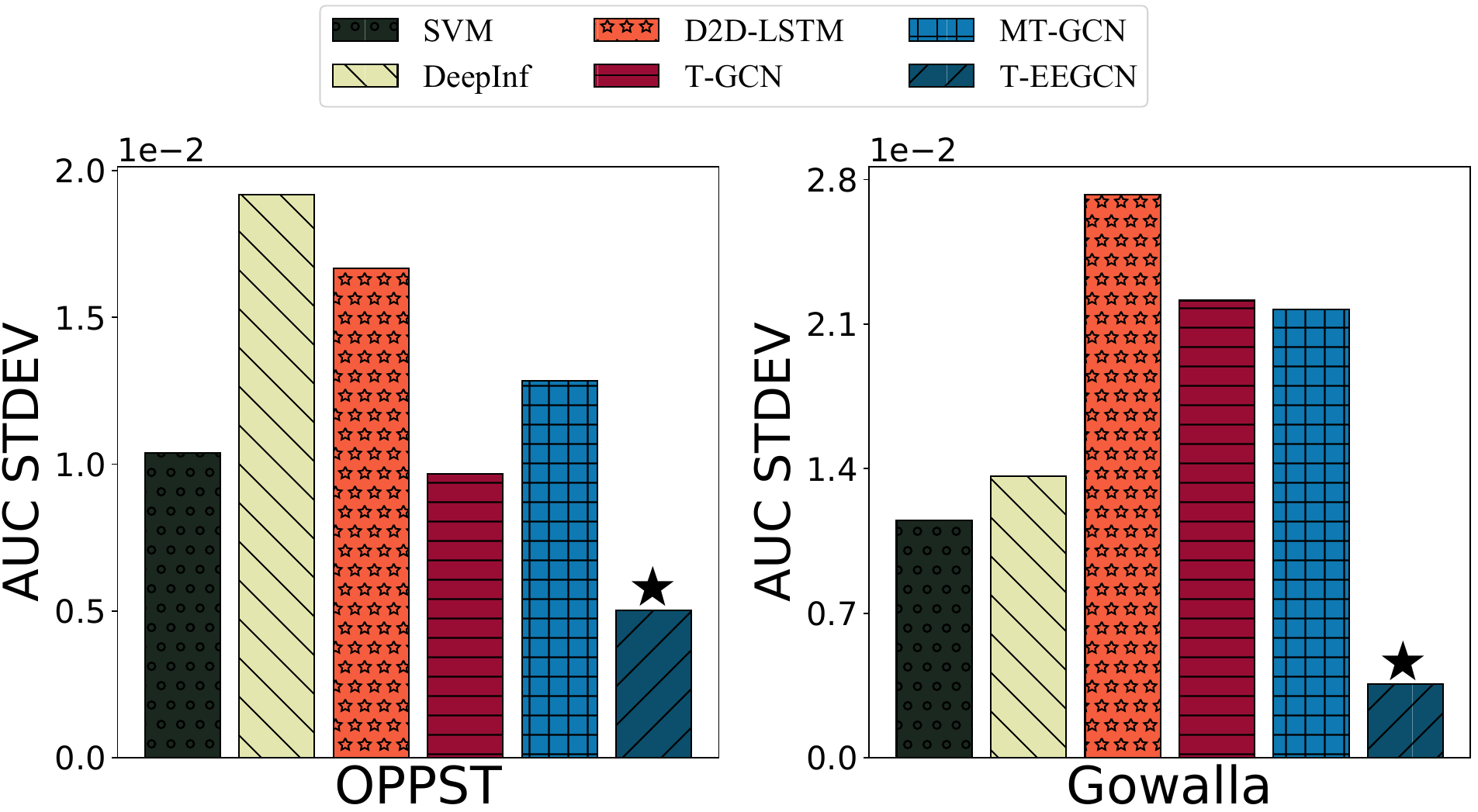}
    \caption{AUC STDEV of Different Approaches on the Datasets OPPST and Gowalla.}
    \label{fig:AUC STDEV}
\end{figure}

\textbf{Content Categories Robustness Analysis.} We calculate the models prediction AUC STDEV (standard deviation, \emph{the smaller, the better}) for different categories of contents. As seen in Fig. \ref{fig:AUC STDEV}, the AUC STDEV of T-EEGCN is much lower than those of baselines in both datasets. The lowest AUC STDEV demonstrates the robustness of T-EEGCN on different categories of content, indicating that it can obtain stable results for different contents, which is not available for other models focusing on limited features.

\subsection{Multi-feature Extraction Performance Analysis}
In this section, we perform ablation studies of T-EEGCN based on the AUC performance on both datasets as a demonstration to illustrate the role of different components and features in implementing T-EEGCN. Table \ref{table:feature ext} demonstrates the multiple ablation variants of T-EEGCN in which various components and features are deactivated.

\begin{table}[ht]
\caption{Feature extraction for different models}
\label{table:feature ext}
\resizebox{8.5cm}{!}{
\centering
\begin{tabular}{@{}cccccccc@{}}
\hline
Mode        & Dynamic        & E.W.          & E.E.         & LSTM       & Social           & Spatio        & Temporal                 \\ \hline
T-EEGCN\#1   & -             & $\checkmark$  & $\checkmark$  & -            & $\checkmark$   & $\checkmark$ & -                \\
T-EEGCN\#2   & $\checkmark$  & -             & $\checkmark$             & $\checkmark$ & $\checkmark$   & -            & $\checkmark$     \\
T-EEGCN\#3   & weak          & $\checkmark$  & -             & $\checkmark$ & $\checkmark$   & weak       & $\checkmark$     \\
T-EEGCN      & enhanced      & $\checkmark$  & $\checkmark$  & $\checkmark$ & $\checkmark$   & enhanced   & $\checkmark$     \\ \hline
\end{tabular}}
\end{table}

\begin{figure}
    \centering
    \includegraphics[width=\columnwidth]{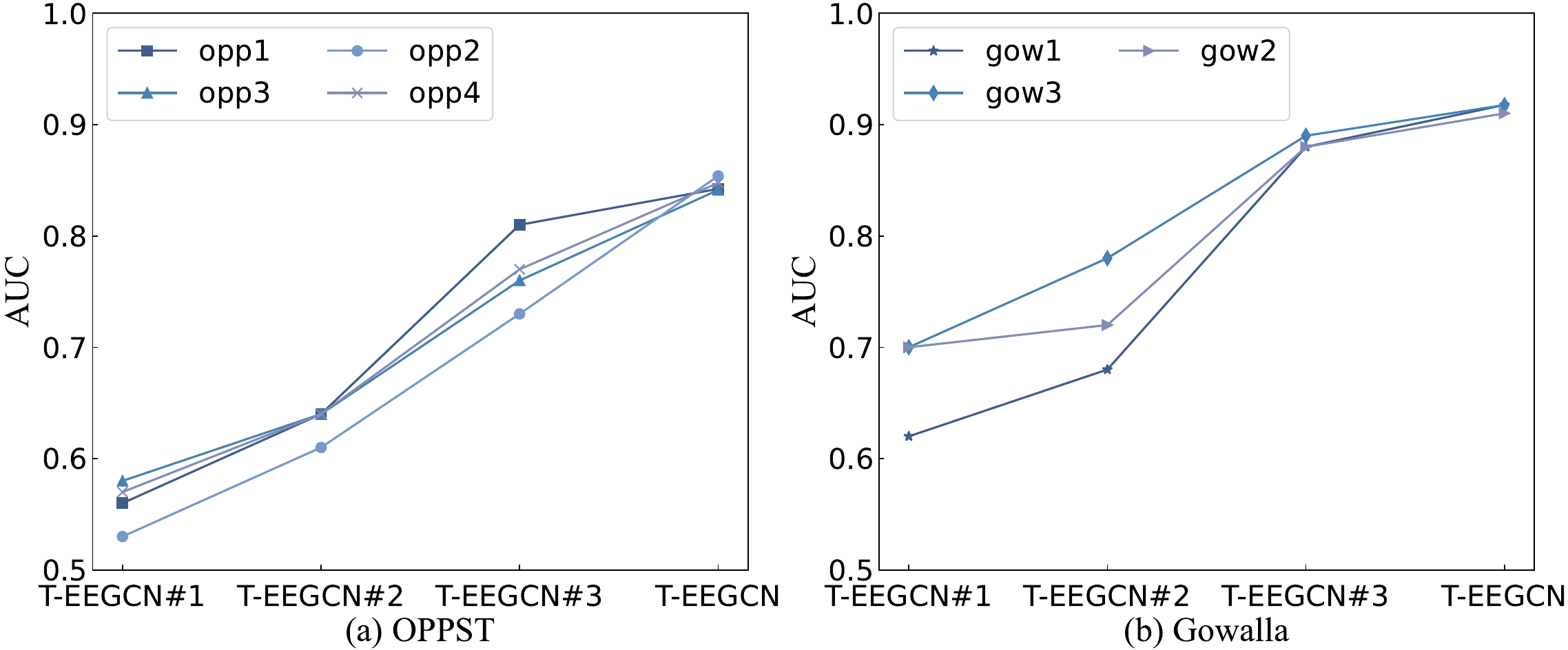}
    \caption{AUC Performance of Ablation Models and T-EEGCN.}
    \label{fig:ablation result}
\end{figure}

\textbf{AUC performance analysis of ablation models.} Fig. \ref{fig:ablation result} shows the AUC performance of ablation models and our T-EEGCN.

\begin{itemize}
    \item The starting points of the curves in Fig. \ref{fig:ablation result} shows the AUC of \textbf{T-EEGCN\#1 (with static graph and deactivate LSTM)}. It can be seen that \#1 performs the worst among all models. Since \#1 updates the node embedding based on the static graph structure, the prediction performance on unknown nodes/graphs is not guaranteed. Besides, \#1 lacks the mechanism to capture temporal features effectively. In contrast, T-EEGCN has a significant increase in performance, indicating that the learning of the \textbf{dynamic graph} structures (adjusting the three learnable parameters based on the dynamic graph structures) and LSTM are key mechanisms to extract \textbf{temporal features}.
    \item \textbf{T-EEGCN\#2} deactivate \textbf{Edge Weighted} from T-EEGCN. Therefore it only considers the social and temporal features and ignores the effect of spatio features on CDSs. The improvement of T-EEGCN compared to \#2 illustrates the necessity to consider \textbf{spatio features.}
    \item \textbf{T-EEGCN\#3 (deactivate Edge Enhanced learning)} is the best among the three ablation models. However, its performance is still worse than T-EEGCN, indicating that exploiting spatio distance as negative weights to adjust node information propagation is not always effective. The additional learnable parameters $W_{\phi}$ and $W_{\theta} $introduced by T-EEGCN enhance the model's capability to extract spatio features and handle dynamic graphs.
\end{itemize}

The ablation studies demonstrate that spatio-temporal and social features play a key role in the prediction (since we cannot deactivate the graph convolution from the T-EEGCN, we analyze the importance of social features by comparing T-EEGCN with D2D-LSTM in Section \ref{er}). In terms of model performance, we can infer that in both OPPST and Gowalla services, the relationship between the influence of features on the prediction model is \textbf{temporal $>$ spatio $>$ social features}. The finding makes sense because our problem is typically a spatio-temporal prediction problem. 

However, we cannot deny the importance of social features. Firstly, T-EEGCN significantly improves prediction performance compared with the D2D-LSTM model that only considers spatio-temporal features (25\%/12\% for average F1/AUC in OPPST and 6\%/4\% for average F1/AUC in Gowalla). Second, the extraction of spatio features by T-EEGCN is built on the extraction of social features. EEGCB trains the learnable parameters and adjusts the node embedding rules by fusing spatio distance and dynamic graph structures. Hence, the structures of CDSs networks (i.e., social features) is indispensable.



\begin{figure}[h]
    \centering
    \includegraphics[width=\columnwidth]{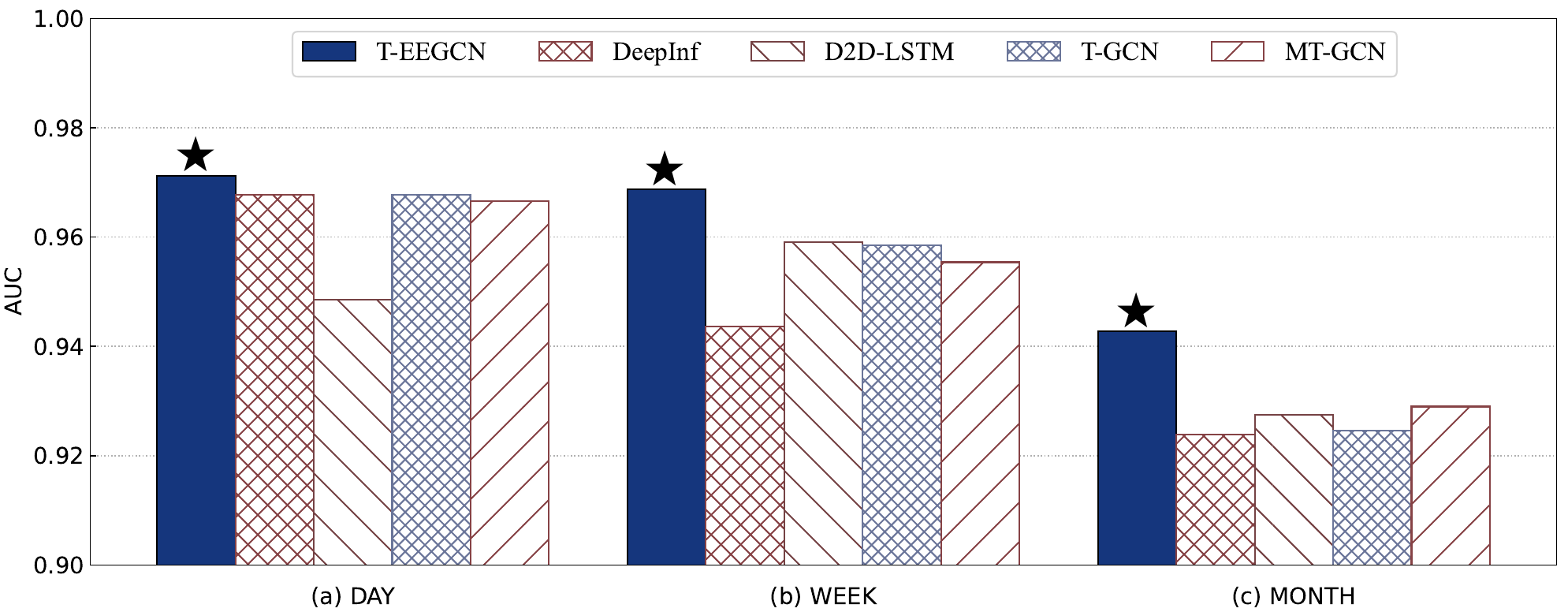}
    \caption{Comparison with Baselines on OPPST Varying the Time Scales}
    \label{fig:OPPST Time scalability}
\end{figure}

\begin{figure}[h]
    \centering
    \includegraphics[width=\columnwidth]{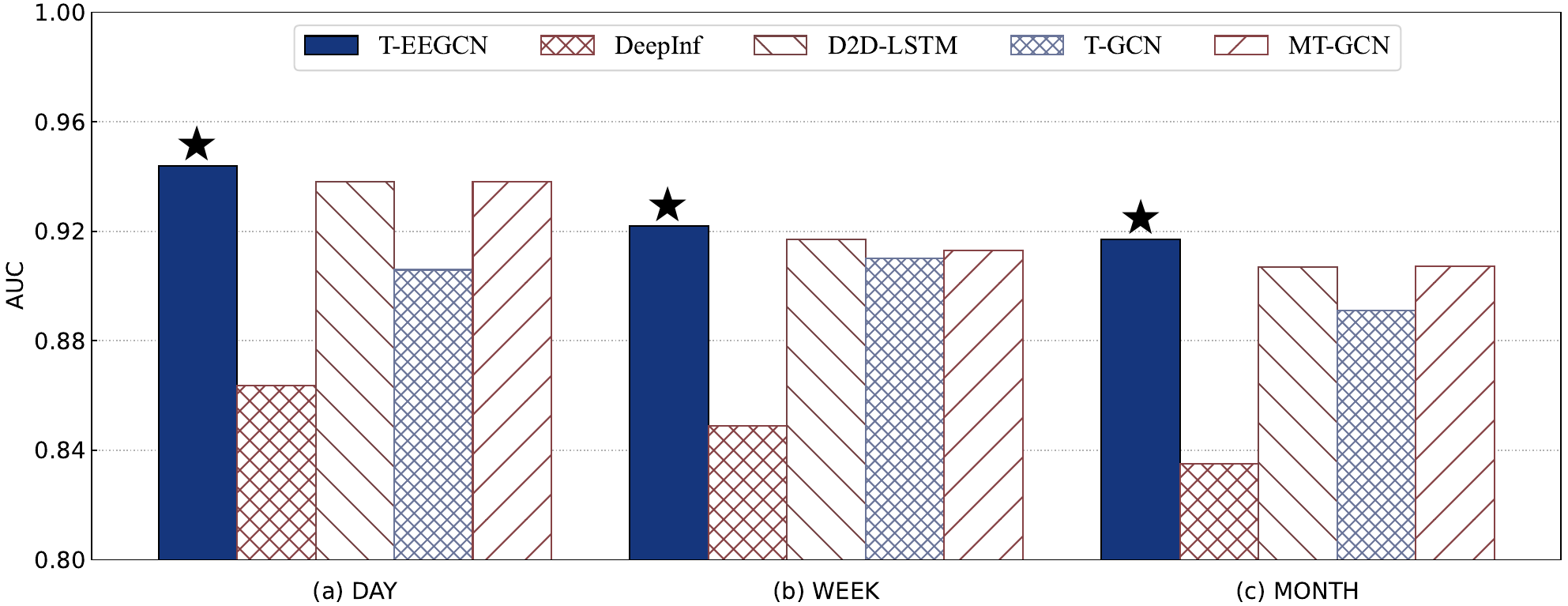}
    \caption{Comparison with Baselines on Gowalla Varying the Time Scales}
    \label{fig:Gowalla Time scalability}
\end{figure}

\subsection{Temporal Scalability Analysis}
\N{Besides the optimization of existing APs (content caching or service orchestration), the growing CDSs often require the frequent deployment of new APs to cope with the tremendous data traffic. Optimization for existing APs is aimed at dissolving the congestion in a short period. However, deploying new APs must consider the potential CDSs for multiple days or even months, requiring network intelligence with strong temporal scalability. Therefore, we perform the temporal scalability test to evaluate the effectiveness of various approaches in handling long-short term predictions.}

\N{Fig. \ref{fig:OPPST Time scalability} and Fig. \ref{fig:Gowalla Time scalability} show the performances with various prediction time windows sizes $T$, and time scales (days in one time window). Specifically, different experiment settings are designed for performance tests on the prediction of CDSs hotspots within one day ($T=6$), one week ($T=3$), and one month ($T=2$). Note that to avoid the bias of categories on the model, we use all categories under the corresponding dataset in the test.}

\N{From Fig. \ref{fig:OPPST Time scalability} and Fig. \ref{fig:Gowalla Time scalability}, we can conclude that:}

\N{\textbf{(i)} As the prediction time scale becomes larger and the time window size becomes shortened, the prediction accuracy of each model decreases. One reason is that as the time scale becomes larger, the number of CDSs also becomes larger, and the CDSs networks become more complex, which will increase the difficulty of GCN encoding. The other reason is that when the time window size decreases, fewer historical features are available to be captured by the LSTM.}

\N{\textbf{(ii)} Although the model suffers from two limitations, T-EEGCN achieves the highest prediction accuracy at all time scales, which proves the effectiveness of the proposed multi-feature capture. The above temporal scalability tests demonstrate that the proposed network intelligence framework can effectively predict the CDSs hotspots for long-short periods, which helps to optimize various services of the 5G UDNs.}

\M{\textbf{(iii)} The proposed approach outperforms baselines and stat-of-the-art on day-granularity predictions, while the model update time on the central cloud is much less than one day. Since we performed experiments on the server with 16 CPUs, 256GB RAM, 2$\times$NVIDIA GeForce RTX 3090, and the training/inference time of the prediction model with day granularity was 2h30min/97ms on average. Thus the proposal guarantees instantaneous at the smallest time granularity currently considered.}

\subsection{Case Study}
We choose the CDSs distribution for the last 30 days (September 2010 to October 2010) in the range of 1216m*1216m (16*16 areas) in New York under Gowalla to plot the CDSs hotspots heat map. Specifically, we calculate the ground truth and the prediction results to form a ground-truth matrix and three prediction matrices (in the form of a probability matrix). The ground truth matrix only has values 0: non-hotspot and 1: hotspot, while the continuous value of the probability matrix represents the predicted hotspot probability.

\begin{figure}[t]
\centering
\subfigure[Ground truth]{\label{a}
\includegraphics[width=4cm]{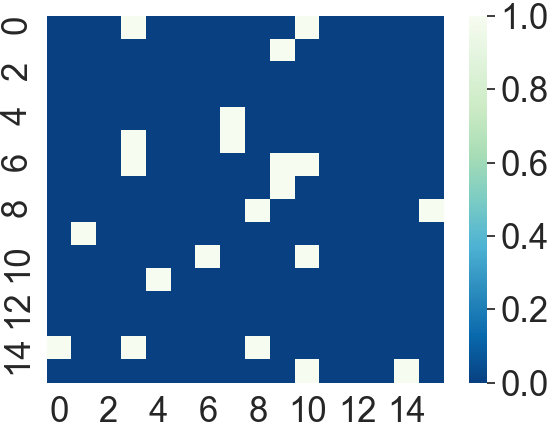}
}
\hspace{-1mm}
\subfigure[MT-GCN]{\label{b}
\includegraphics[width=4cm]{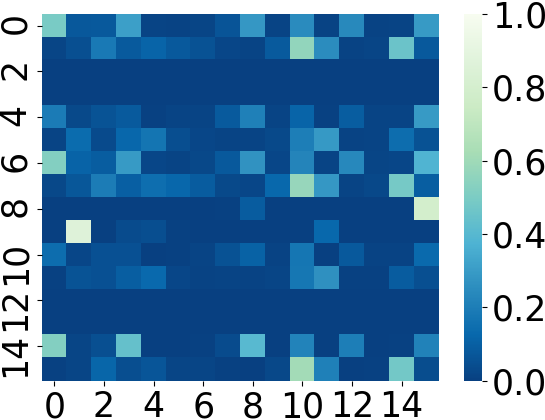}
}
\hspace{-1mm}
\\
\subfigure[D2D-LSTM]{\label{c}
\includegraphics[width=4cm]{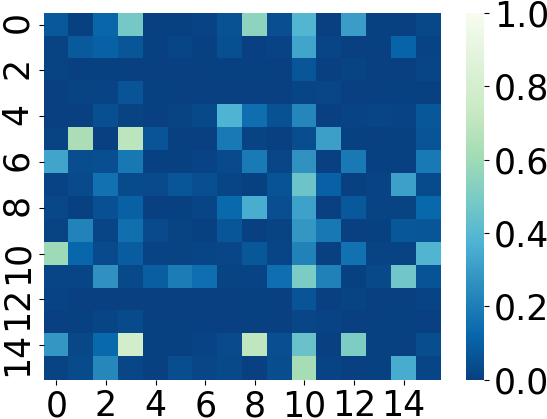}
}
\hspace{-1mm}
\subfigure[T-EEGCN]{\label{d}
\includegraphics[width=4cm]{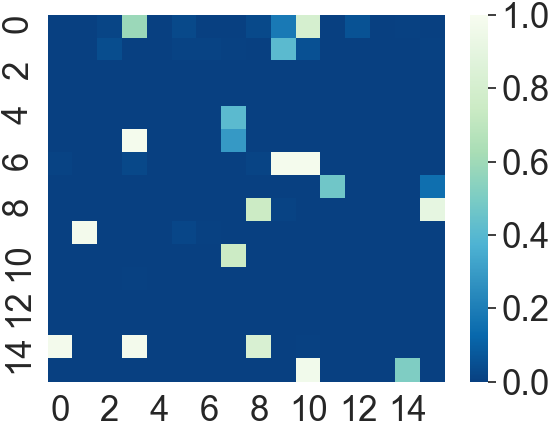}
}
\caption{CDSs heatmaps of ground truth and different models' prediction during Sep. 2010 - Oct. 2010}
\label{fig:heatmap}
\end{figure}

The results are reported in Fig. \ref{fig:heatmap}. In the heatmap, blue represents the hotspot's negative instance, while white represents the positive instance. Moreover, the color depth indicates the hotspot probability, in which the color close to dark blue indicates low probability while that close to white means high likelihood.

As illustrated in Fig. \ref{fig:heatmap}, the two baselines are deficient in terms of prediction accuracy (predicted hotspots distribution does not match the real distribution) and hotspots discovery capability (low confidence in discovering true hotspots with shallow whites). In contrast, our framework can provide predictions consistent with the distribution of real CDSs and possesses a very high confidence level for true hotspots, which are sufficient for guiding real-world decisions.

\section{Conclusion}\label{conclusion}
\N{We study the spatio fine-grained and generalized CDSs hotspots prediction in 5G UDNs. We first adopt the Geohash encoding and propose a spatio-temporal subgraph sampling algorithm for fine-grained and arbitrary areas CDSs modeling. Then, we propose a novel multi-feature extraction framework T-EEGCN, in which we propose a novel edge-enhanced graph convolution block to accurately encode dynamic CDSs networks by considering spatio-social features. Besides, we adopt the LSTM to capture the temporal dependency. The analysis and experiments on two real-world CDSs datasets verify the influence of spatio-temporal-social features and the effectiveness of T-EEGCN.}

\section*{Acknowledgement}
This work is partially supported by the National Key Research and Development Program of China (Grant No. 2019YFB2101901), the National Science Foundation of China (Grant No.62072332), and Australian Research Council Linkage Project (LP180100750).


%

\appendices




\ifCLASSOPTIONcaptionsoff
  \newpage
\fi



%



\bibliographystyle{IEEEtran}
\bibliography{manuscript}

\begin{thebibliography}{10}
\providecommand{\url}[1]{#1}
\csname url@samestyle\endcsname
\providecommand{\newblock}{\relax}
\providecommand{\bibinfo}[2]{#2}
\providecommand{\BIBentrySTDinterwordspacing}{\spaceskip=0pt\relax}
\providecommand{\BIBentryALTinterwordstretchfactor}{4}
\providecommand{\BIBentryALTinterwordspacing}{\spaceskip=\fontdimen2\font plus
\BIBentryALTinterwordstretchfactor\fontdimen3\font minus
  \fontdimen4\font\relax}
\providecommand{\BIBforeignlanguage}[2]{{%
\expandafter\ifx\csname l@#1\endcsname\relax
\typeout{** WARNING: IEEEtran.bst: No hyphenation pattern has been}%
\typeout{** loaded for the language `#1'. Using the pattern for}%
\typeout{** the default language instead.}%
\else
\language=\csname l@#1\endcsname
\fi
#2}}
\providecommand{\BIBdecl}{\relax}
\BIBdecl

\bibitem{9817385}
X.~Chen, Q.~Xiang, L.~Kong, H.~Xu, and X.~Liu, ``Learning from fm
  communications: Toward accurate, efficient, all-terrain vehicle
  localization,'' \emph{IEEE/ACM Transactions on Networking}, pp. 1--16, 2022.

\bibitem{7839234}
F.-H. Tseng, Y.-M. Jheng, L.-D. Chou, H.-C. Chao, and V.~C. Leung, ``Link-aware
  virtual machine placement for cloud services based on service-oriented
  architecture,'' \emph{IEEE Transactions on Cloud Computing}, vol.~8, no.~4,
  pp. 989--1002, 2020.

\bibitem{7539325}
J.~Qiao, Y.~He, and X.~S. Shen, ``Proactive caching for mobile video streaming
  in millimeter wave 5g networks,'' \emph{IEEE Transactions on Wireless
  Communications}, vol.~15, no.~10, pp. 7187--7198, 2016.

\bibitem{8865093}
J.~Park, S.~Samarakoon, M.~Bennis, and M.~Debbah, ``Wireless network
  intelligence at the edge,'' \emph{Proceedings of the IEEE}, vol. 107, no.~11,
  pp. 2204--2239, 2019.

\bibitem{7476821}
M.~{Kamel}, W.~{Hamouda}, and A.~{Youssef}, ``Ultra-dense networks: A survey,''
  \emph{IEEE Communications Surveys Tutorials}, vol.~18, no.~4, 2016.

\bibitem{8030322}
N.~Abbas, Y.~Zhang, A.~Taherkordi, and T.~Skeie, ``Mobile edge computing: A
  survey,'' \emph{IEEE Internet of Things Journal}, vol.~5, no.~1, pp.
  450--465, 2018.

\bibitem{10.1145/3209582.3209606}
C.~Zhang and P.~Patras, ``Long-term mobile traffic forecasting using deep
  spatio-temporal neural networks,'' in \emph{Proceedings of the Eighteenth ACM
  International Symposium on Mobile Ad Hoc Networking and Computing}.\hskip 1em
  plus 0.5em minus 0.4em\relax Association for Computing Machinery, 2018.

\bibitem{8845204}
E.~{Mededovic}, V.~G. {Douros}, and P.~{Mähönen}, ``Node centrality metrics
  for hotspots analysis in telecom big data,'' in \emph{INFOCOM WKSHPS}, 2019.

\bibitem{article}
H.~Zhang, X.~Wang, J.~Chen, C.~Wang, and J.~Li, ``D2d-lstm: Lstm-based path
  prediction of content diffusion tree in device-to-device social networks,''
  \emph{AAAI}, vol.~34, 2020.

\bibitem{wang2015characterizing}
H.~Wang, J.~Ding, Y.~Li, P.~Hui, J.~Yuan, and D.~Jin, ``Characterizing the
  spatio-temporal inhomogeneity of mobile traffic in large-scale cellular data
  networks,'' in \emph{Proceedings of the 7th International Workshop on Hot
  Topics in Planet-scale mObile computing and online Social neTworking}, 2015,
  pp. 19--24.

\bibitem{8117559}
X.~{Wang}, Z.~{Zhou}, Z.~{Yang}, Y.~{Liu}, and C.~{Peng}, ``Spatio-temporal
  analysis and prediction of cellular traffic in metropolis,'' in \emph{2017
  IEEE 25th International Conference on Network Protocols (ICNP)}, 2017.

\bibitem{8737488}
D.~{Bega}, M.~{Gramaglia}, M.~{Fiore}, A.~{Banchs}, and X.~{Costa-Perez},
  ``Deepcog: Cognitive network management in sliced 5g networks with deep
  learning,'' in \emph{IEEE INFOCOM 2019 - IEEE Conference on Computer
  Communications}, 2019.

\bibitem{8057090}
J.~{Wang}, J.~{Tang}, Z.~{Xu}, Y.~{Wang}, G.~{Xue}, X.~{Zhang}, and D.~{Yang},
  ``Spatiotemporal modeling and prediction in cellular networks: A big data
  enabled deep learning approach,'' in \emph{IEEE INFOCOM 2017 - IEEE
  Conference on Computer Communications}, 2017.

\bibitem{8292737}
C.~{Huang}, C.~{Chiang}, and Q.~{Li}, ``A study of deep learning networks on
  mobile traffic forecasting,'' in \emph{2017 IEEE 28th Annual International
  Symposium on Personal, Indoor, and Mobile Radio Communications (PIMRC)},
  2017.

\bibitem{7890496}
R.~{Li}, Z.~{Zhao}, J.~{Zheng}, C.~{Mei}, Y.~{Cai}, and H.~{Zhang}, ``The
  learning and prediction of application-level traffic data in cellular
  networks,'' \emph{IEEE Transactions on Wireless Communications}, vol.~16,
  no.~6, 2017.

\bibitem{koren2022advances}
Y.~Koren, S.~Rendle, and R.~Bell, ``Advances in collaborative filtering,''
  \emph{Recommender systems handbook}, pp. 91--142, 2022.

\bibitem{lu2021time}
Y.~Lu, Y.~He, Y.~Cai, Z.~Peng, and Y.~Tang, ``Time-aware neural collaborative
  filtering with multi-dimensional features on academic paper recommendation,''
  in \emph{2021 IEEE 24th International Conference on Computer Supported
  Cooperative Work in Design (CSCWD)}.\hskip 1em plus 0.5em minus 0.4em\relax
  IEEE, 2021, pp. 1052--1057.

\bibitem{Zeng2021PRRCUCAP}
J.~Zeng, H.~Tang, Y.~Zhao, M.~Gao, and J.~Wen, ``Pr-rcuc: A poi recommendation
  model using region-based collaborative filtering and user-based mobile
  context,'' \emph{Mob. Networks Appl.}, vol.~26, pp. 2434--2444, 2021.

\bibitem{wei2006time}
W.~W. Wei, ``Time series analysis,'' in \emph{The Oxford Handbook of
  Quantitative Methods in Psychology: Vol. 2}, 2006.

\bibitem{ABBASI202119}
M.~Abbasi, A.~Shahraki, and A.~Taherkordi, ``Deep learning for network traffic
  monitoring and analysis (ntma): A survey,'' \emph{Computer Communications},
  vol. 170, pp. 19--41, 2021.

\bibitem{10.1007/978-3-030-33778-0_11}
A.~Azari, P.~Papapetrou, S.~Denic, and G.~Peters, ``Cellular traffic prediction
  and classification: A comparative evaluation of lstm and arima,'' in
  \emph{Discovery Science: 22nd International Conference, DS 2019, Split,
  Croatia, October 28–30, 2019, Proceedings}.\hskip 1em plus 0.5em minus
  0.4em\relax Berlin, Heidelberg: Springer-Verlag, 2019, p. 129–144.

\bibitem{zhang2019deep}
C.~Zhang, H.~Zhang, J.~Qiao, D.~Yuan, and M.~Zhang, ``Deep transfer learning
  for intelligent cellular traffic prediction based on cross-domain big data,''
  \emph{IEEE Journal on Selected Areas in Communications}, vol.~37, no.~6, pp.
  1389--1401, 2019.

\bibitem{zhou2022large}
X.~Zhou, Y.~Zhang, Z.~Li, X.~Wang, J.~Zhao, and Z.~Zhang, ``Large-scale
  cellular traffic prediction based on graph convolutional networks with
  transfer learning,'' \emph{Neural Computing and Applications}, vol.~34,
  no.~7, pp. 5549--5559, 2022.

\bibitem{9184280}
K.~He, X.~Chen, Q.~Wu, S.~Yu, and Z.~Zhou, ``Graph attention spatial-temporal
  network with collaborative global-local learning for citywide mobile traffic
  prediction,'' \emph{IEEE Transactions on Mobile Computing}, vol.~21, no.~4,
  pp. 1244--1256, 2022.

\bibitem{zhang2021dual}
C.~Zhang, S.~Dang, B.~Shihada, and M.-S. Alouini, ``Dual attention-based
  federated learning for wireless traffic prediction,'' in \emph{IEEE INFOCOM
  2021-IEEE conference on computer communications}.\hskip 1em plus 0.5em minus
  0.4em\relax IEEE, 2021, pp. 1--10.

\bibitem{shafiq2011characterizing}
M.~Z. Shafiq, L.~Ji, A.~X. Liu, and J.~Wang, ``Characterizing and modeling
  internet traffic dynamics of cellular devices,'' \emph{ACM SIGMETRICS
  Performance Evaluation Review}, vol.~39, no.~1, pp. 265--276, 2011.

\bibitem{6757900}
D.~Lee, S.~Zhou, X.~Zhong, Z.~Niu, X.~Zhou, and H.~Zhang, ``Spatial modeling of
  the traffic density in cellular networks,'' \emph{IEEE Wireless
  Communications}, vol.~21, no.~1, pp. 80--88, 2014.

\bibitem{1203886}
Y.~Shu, M.~Yu, J.~Liu, and O.~Yang, ``Wireless traffic modeling and prediction
  using seasonal arima models,'' in \emph{IEEE International Conference on
  Communications, 2003. ICC '03.}, vol.~3, 2003, pp. 1675--1679 vol.3.

\bibitem{DUAN2022109156}
J.-H. Duan, W.~Li, X.~Zhang, and S.~Lu, ``Forecasting fine-grained city-scale
  cellular traffic with sparse crowdsourced measurements,'' \emph{Computer
  Networks}, vol. 214, p. 109156, 2022.

\bibitem{9751165}
E.~Tuna and A.~Soysal, ``Multivariate spatio-temporal cellular traffic
  prediction with handover based clustering,'' in \emph{2022 56th Annual
  Conference on Information Sciences and Systems (CISS)}, 2022, pp. 55--59.

\bibitem{9112663}
L.~Yu, M.~Li, W.~Jin, Y.~Guo, Q.~Wang, F.~Yan, and P.~Li, ``Step: A
  spatio-temporal fine-granular user traffic prediction system for cellular
  networks,'' \emph{IEEE Transactions on Mobile Computing}, vol.~20, no.~12,
  pp. 3453--3466, 2021.

\bibitem{9127090}
J.~Chu, X.~Wang, K.~Qian, L.~Yao, F.~Xiao, J.~Li, and Z.~Yang, ``Passenger
  demand prediction with cellular footprints,'' \emph{IEEE Transactions on
  Mobile Computing}, vol.~21, no.~1, pp. 252--263, 2022.

\bibitem{8903457}
Q.~Hu, S.~Wang, X.~Cheng, J.~Zhang, and W.~Lv, ``Cost-efficient mobile
  crowdsensing with spatial-temporal awareness,'' \emph{IEEE Transactions on
  Mobile Computing}, vol.~20, no.~3, pp. 928--938, 2021.

\bibitem{9165195}
H.~Li, F.~Lin, X.~Lu, C.~Xu, G.~Huang, J.~Zhang, Q.~Mei, and X.~Liu,
  ``Systematic analysis of fine-grained mobility prediction with on-device
  contextual data,'' \emph{IEEE Transactions on Mobile Computing}, vol.~21,
  no.~3, pp. 1096--1109, 2022.

\bibitem{kipf2017semisupervised}
T.~N. Kipf and M.~Welling, ``Semi-supervised classification with graph
  convolutional networks,'' 2017.

\bibitem{Bing2018Spatio}
B.~Yu, H.~Yin, and Z.~Zhu, ``Spatio-temporal graph convolutional networks: A
  deep learning framework for traffic forecasting,'' in \emph{IJCAI}, 2018.

\bibitem{ijcai2020-326}
R.~Huang, C.~Huang, Y.~Liu, G.~Dai, and W.~Kong, ``Lsgcn: Long short-term
  traffic prediction with graph convolutional networks,'' in \emph{Proceedings
  of the Twenty-Ninth International Joint Conference on Artificial
  Intelligence, {IJCAI-20}}.\hskip 1em plus 0.5em minus 0.4em\relax
  International Joint Conferences on Artificial Intelligence Organization, 7
  2020.

\bibitem{qiu2018deepinf}
J.~Qiu, J.~Tang, H.~Ma, Y.~Dong, K.~Wang, and J.~Tang, ``Deepinf: Social
  influence prediction with deep learning,'' in \emph{ACM SIGKDD}, 2018.

\bibitem{2019Social2}
Y.~Xiao, C.~Song, and Y.~Liu, ``Social hotspot propagation dynamics model based
  on multidimensional attributes and evolutionary games,'' \emph{Communications
  in Nonlinear Science and Numerical Simulation}, vol.~67, no. FEB., pp.
  13--25, 2019.

\bibitem{Liu2019}
C.~H. Liu, J.~Xu, J.~Tang, and J.~Crowcroft, ``{Social-Aware Sequential
  Modeling of User Interests: A Deep Learning Approach},'' \emph{IEEE
  Transactions on Knowledge and Data Engineering}, vol.~31, no.~11, pp.
  2200--2212, 2019.

\bibitem{scellato2011socio}
S.~Scellato, A.~Noulas, R.~Lambiotte, and C.~Mascolo, ``Socio-spatial
  properties of online location-based social networks.'' \emph{ICWSM}, vol.~11,
  pp. 329--336, 2011.

\bibitem{10.1162/neco_a_01199}
\BIBentryALTinterwordspacing
Y.~Yu, X.~Si, C.~Hu, and J.~Zhang, ``{A Review of Recurrent Neural Networks:
  LSTM Cells and Network Architectures},'' \emph{Neural Computation}, vol.~31,
  no.~7, pp. 1235--1270, 07 2019. [Online]. Available:
  \url{https://doi.org/10.1162/neco\_a\_01199}
\BIBentrySTDinterwordspacing

\bibitem{platt1999probabilistic}
J.~Platt \emph{et~al.}, ``Probabilistic outputs for support vector machines and
  comparisons to regularized likelihood methods,'' \emph{Advances in large
  margin classifiers}, 1999.

\bibitem{Zhao2020}
L.~Zhao, Y.~Song, C.~Zhang, Y.~Liu, P.~Wang, T.~Lin, M.~Deng, and H.~Li,
  ``{T-GCN: A Temporal Graph Convolutional Network for Traffic Prediction},''
  \emph{IEEE Transactions on Intelligent Transportation Systems}, vol.~21,
  no.~9, pp. 3848--3858, 2020.

\bibitem{liu2019characterizing}
Y.~Liu, X.~Shi, L.~Pierce, and X.~Ren, ``Characterizing and forecasting user
  engagement with in-app action graph: A case study of snapchat,'' in \emph{ACM
  SIGKDD}, 2019.

\end{thebibliography}
%

\vspace{-15mm}

\begin{IEEEbiography}[{\includegraphics[width=1in,height=1.2in,clip,keepaspectratio]{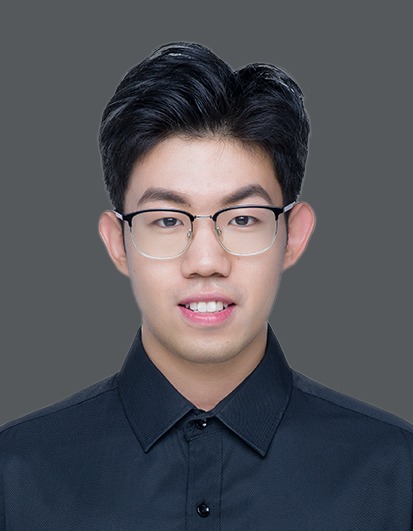}}]{Shaoyuan Huang}
Shaoyuan Huang received a B.S. degree from Tianjin University, Tianjin, China, in 2020. He is currently pursuing an M.S. degree in the College of Intelligence and Computing, Tianjin University. His current research interests include spatial-temporal prediction, recommend systems and graph convolution networks.
\end{IEEEbiography}

\vspace{-15mm}

\begin{IEEEbiography}[{\includegraphics[width=1in,height=1.2in,clip,keepaspectratio]{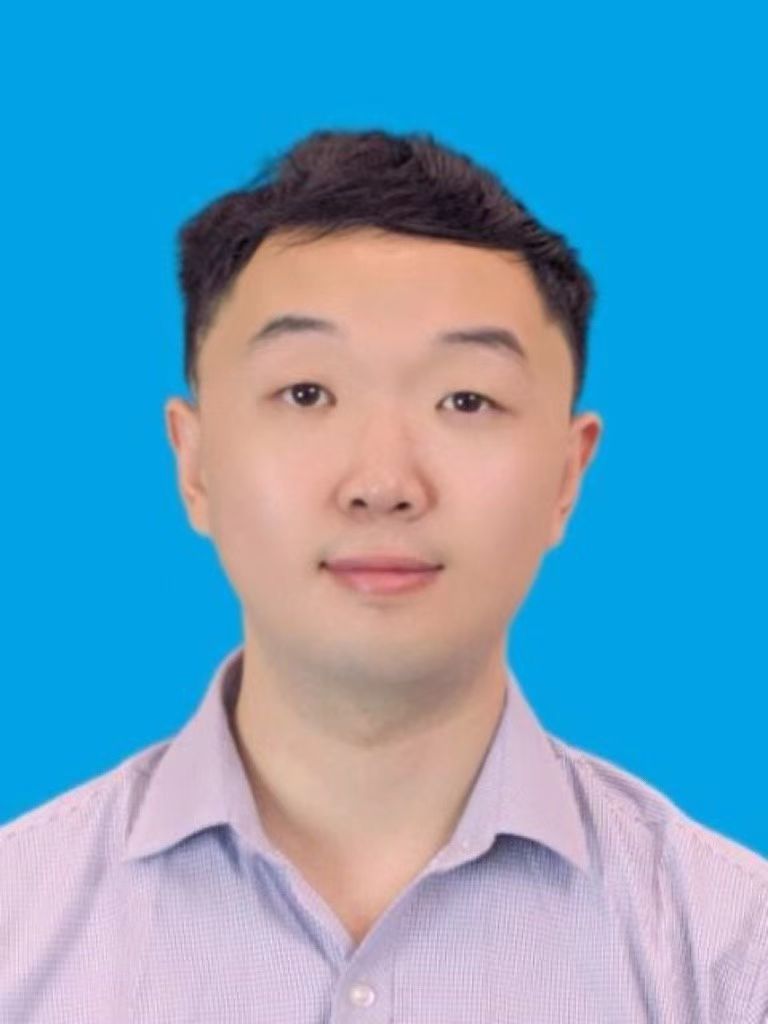}}]{Heng Zhang}
Heng Zhang is currently pursuing a PhD degree from the School of Computer Science and Technology, College of Intelligence and Computing, Tianjin University, Tianjin, China. His current research interests include D2D content propagation, recommend system and edge computing
\end{IEEEbiography}

\vspace{-15mm}

\begin{IEEEbiography}[{\includegraphics[width=1in,height=1.2in,clip,keepaspectratio]{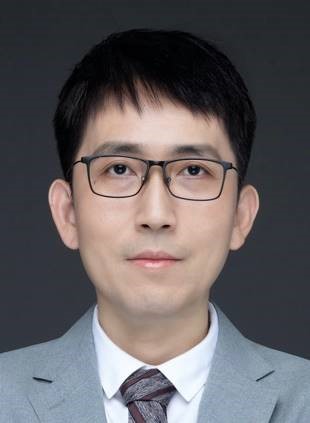}}]{Xiaofei Wang} [S'06, M'13, SM'18] (Senior Member, IEEE) received the B.S. degree from Huazhong University of Science and Technology, China, and received M.S. and Ph.D. degrees from Seoul National University, Seoul, South Korea. He was a Postdoctoral Fellow with The University of British Columbia, Vancouver, Canada, from 2014 to 2016. He is currently a Professor with the Tianjin Key Laboratory of Advanced Networking, College of Intelligence and Computing, Tianjin University, Tianjin, China. Focusing on the research of edge computing, edge intelligence, and edge systems, he has published more than 150 technical papers in IEEE JSAC, TCC, ToN, TWC, IoTJ, COMST, TMM, INFOCOM, ICDCS and so on. In 2017, he was the recipient of the "IEEE ComSoc Fred W. Ellersick Prize", and in 2022, he received the "IEEE ComSoc Asia-Pacific Outstanding Paper Award".
\end{IEEEbiography}

\vspace{-15mm}

\begin{IEEEbiography}[{\includegraphics[width=1in,height=1.2in,clip,keepaspectratio]{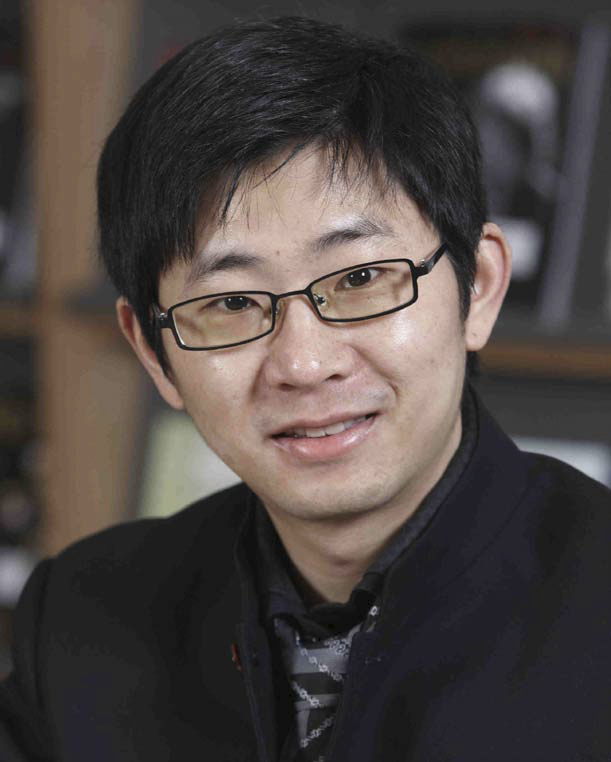}}]{Min Chen} (Fellow, IEEE) has been a Full Professor with the School of Computer Science and Technology, Huazhong University of Science and Technology (HUST), since February 2012. He is the Director of the Embedded and Pervasive Computing Laboratory and the Director of the Data Engineering Institute, HUST. Before he joined HUST, he was an Assistant Professor with the School of Computer Science and Engineering, Seoul National University. His Google Scholar citations reached over 34,450 with an H-index of 89. His top paper was cited 3,825 times. He was selected as a Highly Cited Researcher from 2018 to 2021. He received the IEEE Communications Society Fred W. Ellersick Prize in 2017 and the IEEE Jack Neubauer Memorial Award in 2019. He is the Chair of the IEEE Globecom 2022 eHealth Symposium. He is the Founding Chair of the IEEE Computer Society Special Technical Communities on Big Data.
\end{IEEEbiography}

\vspace{-15mm}

\begin{IEEEbiography}[{\includegraphics[width=1in,height=1.2in,clip,keepaspectratio]{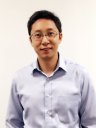}}]{Jianxin Li} received a Ph.D. degree in computer  science  from  the  Swinburne  University  of Technology, Melbourne, VIC, Australia, in 2009.He  is  an  Associate  Professor  with  the  School of Information Technology, Deakin University, Burwood, VIC, Australia. His current research interests include database query processing and optimization,  social  network  analytics,  and  traffic network data processing.
\end{IEEEbiography}

\vspace{-15mm}

\begin{IEEEbiography}[{\includegraphics[width=1in,height=1.2in,clip,keepaspectratio]{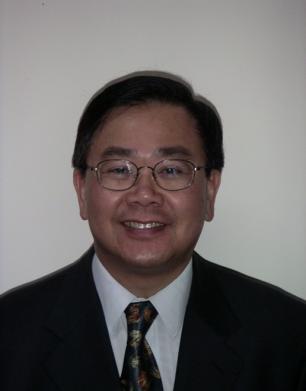}}]{Victor C. M. Leung} (Life  Fellow,  IEEE)  is  a  Distinguished  Professor  of  Computer  Science and  Software  Engineering  with  Shenzhen  University,  Shenzhen,  China.  He  is  also  an  Emeritus Professor of Electrical and Computer Engineering and the Director of the Laboratory for Wireless Networks  and  Mobile  Systems,  University  of  British  Columbia,  Vancouver,  BC,  Canada.  His research is in the broad areas of wireless networks and mobile systems. Dr. Leung  received  the  IEEE  Vancouver  Section  Centennial  Award,  the  2011  UBC  Killam Research Prize, the 2017 Canadian Award for Telecommunications Research, and the 2018 IEEETCGCC  Distinguished  Technical  Achievement  Recognition  Award.He  is  named  in  the  current Clarivate Analytics list of Highly Cited Researchers. He is serving on the editorial boards of IEEE TRANSACTIONS ON GREEN COMMUNICATIONS AND NETWORKING, IEEE TRANSACTIONS ON CLOUD COMPUTING, IEEE NETWORK, and several other journals. He coauthored papers that won the 2017 IEEE ComSoc Fred W. Ellersick Prize, the 2017 IEEE Systems JournalBest Paper Award, the 2018 IEEE CSIM Best Journal Paper Award, and the 2019 IEEE TCGCC Best Journal Paper Award.He is a Fellow of the Royal Society of Canada, Canadian Academy of Engineering, and Engineering Institute of Canada.

\end{IEEEbiography}




\end{document}